\documentclass[nohyper,12pt,letterpaper]{JHEP3}
\usepackage{epsfig}
\usepackage[latin1]{inputenc}
\usepackage{bbm,amsfonts}
\usepackage{graphicx}
\usepackage{amssymb,amsmath}

\author{Marco S. Bianchi$^\ast$,
  Matias Leoni$^{\dag,\hash}$,
  Andrea Mauri$^{\dag,\hash}$,
  Silvia Penati$^\ast$,
  CarloAlberto Ratti$^\ast$
  and Alberto Santambrogio$^\hash$\\\\
  $^\ast$Dipartimento di Fisica, Universit\`a di Milano--Bicocca and
  INFN, Sezione di Milano--Bicocca, Piazza della Scienza 3, I-20126 Milano, Italy \\\\
  $^\dag$Dipartimento di Fisica dell'Universit\`a degli studi di Milano\\\\
  $^\hash$ INFN, Sezione di Milano, via Celoria 16, I-20133 Milano, Italy
  \qquad\\\\
  E-mail: \email{marco.bianchi@mib.infn.it, matias.leoni@mi.infn.it,
    andrea.mauri@mi.infn.it, silvia.penati@mib.infn.it,
    carloalberto.ratti@mib.infn.it, alberto.santambrogio@mi.infn.it }}

\abstract{We study $n$--point correlation functions for chiral primary
  operators in three dimensional supersymmetric Chern--Simons matter
  theories. Our analysis is carried on in ${\cal N}=2$ superspace and
  covers ${\cal N}=2,3$ supersymmetric CFT's, the ${\cal N}=6$ ABJM
  and the ${\cal N}=8$ BLG models. In the limit where the positions of adjacent operators become
  light--like, we find that the one--loop $n$--point correlator
  divided by its tree level expression coincides with a light--like
  $n$--polygon Wilson loop. Remarkably, the result can be simply
  expressed as a linear combination of five dimensional two--mass easy
  boxes.  We manage to evaluate the integrals analytically and find a
  vanishing result, in agreement with previous findings for Wilson
  loops.  }

\preprint{March 2011\\ IFUM-976-FT}

\title{FROM CORRELATORS TO WILSON LOOPS IN CHERN--SIMONS MATTER THEORIES }


\csname @addtoreset\endcsname{equation}{section}

\def\bseq{\begin{subequation}}  
\def\eseq{\end{subequation}}
\def\bsea{\begin{subeqnarray}}  
\def\esea{\end{subeqnarray}}

\newcommand{\beq}{\begin{equation}}
\newcommand{\bea}{\begin{eqnarray}}
\newcommand{\eea}{\end{eqnarray}}
\newcommand{\eeq}{\end{equation}}

\newcommand {\non}{\nonumber}

\newcommand{\Ab}{\bar{A}}

\renewcommand{\a}{\alpha}
\renewcommand{\b}{\beta}

\renewcommand{\d}{\delta}
\newcommand{\pa}{\partial}
\newcommand{\g}{\gamma}
\newcommand{\G}{\Gamma}

\newcommand{\D}{\Delta}
\newcommand{\e}{\epsilon}

\renewcommand{\L}{\Lambda}
\newcommand{\m}{\mu}

\newcommand{\n}{\nu}

\newcommand{\f}{\phi}

\newcommand{\p}{\pi}

\newcommand{\s}{\sigma}

\renewcommand{\t}{\tau}

\newcommand{\Db}{\overline{D}}

\newcommand{\adot}{\dot{\alpha}}

\newcommand{\thb}{\overline{\theta}}

\renewcommand{\thb}{\overline{\theta}}

\def\Mb{\kern 2pt\mathchoice
        {
         \vbox{\hrule width10pt height 0.4pt depth 0pt
         \kern 1.2pt\hbox{\kern -2pt$\displaystyle M$}}}
        {
         \vbox{\hrule width10pt height 0.4pt depth 0pt
         \kern 1.2pt\hbox{\kern -2pt$\textstyle M$}}}
        {
\vbox{\hrule width6pt height 0.4pt depth 0pt
         \kern 1.0pt\hbox{\kern -2pt$\scriptstyle M$}}}
        {
         \vbox{\hrule width5pt height 0.4pt depth 0pt
         \kern 0.8pt\hbox{\kern -2pt$\scriptscriptstyle M$}}}}

\def\Sb{\kern 2pt\mathchoice
        {
         \vbox{\hrule width6pt height 0.4pt depth 0pt
         \kern 1.2pt\hbox{\kern -2pt$\displaystyle S$}}}
        {
         \vbox{\hrule width6pt height 0.4pt depth 0pt
         \kern 1.2pt\hbox{\kern -2pt$\textstyle S$}}}
        {
         \vbox{\hrule width3.5pt height 0.4pt depth 0pt
         \kern 1.0pt\hbox{\kern -2pt$\scriptstyle S$}}}
        {
         \vbox{\hrule width3pt height 0.4pt depth 0pt
         \kern 0.8pt\hbox{\kern -2pt$\scriptscriptstyle S$}}}}

\def\Rb{\kern 2pt\mathchoice
        {
         \vbox{\hrule width5.5pt height 0.4pt depth 0pt
         \kern 1.2pt\hbox{\kern -2.5pt$\displaystyle R$}}}
        {
         \vbox{\hrule width5.5pt height 0.4pt depth 0pt
         \kern 1.2pt\hbox{\kern -2.5pt$\textstyle R$}}}
        {
         \vbox{\hrule width3.5pt height 0.4pt depth 0pt
         \kern 1.0pt\hbox{\kern -2.2pt$\scriptstyle R$}}}
        {
         \vbox{\hrule width3pt height 0.4pt depth 0pt
         \kern 0.8pt\hbox{\kern -2.2pt$\scriptscriptstyle R$}}}}

  \def\pp{{\mathchoice
      {
          \kern 1pt%
          \raise 1pt
          \vbox{\hrule width5pt height0.4pt depth0pt
            \kern -2pt
            \hbox{\kern 2.3pt
              \vrule width0.4pt height6pt depth0pt
              }
            \kern -2pt
            \hrule width5pt height0.4pt depth0pt}%
            \kern 1pt
       }
        {
          \kern 1pt%
          \raise 1pt
          \vbox{\hrule width4.3pt height0.4pt depth0pt
            \kern -1.8pt
            \hbox{\kern 1.95pt
              \vrule width0.4pt height5.4pt depth0pt
              }
            \kern -1.8pt
            \hrule width4.3pt height0.4pt depth0pt}%
            \kern 1pt
        }
        {
          \kern 0.5pt%
          \raise 1pt
          \vbox{\hrule width4.0pt height0.3pt depth0pt
            \kern -1.9pt  
            \hbox{\kern 1.85pt
              \vrule width0.3pt height5.7pt depth0pt
              }
            \kern -1.9pt
            \hrule width4.0pt height0.3pt depth0pt}%
            \kern 0.5pt
        }
        {
          \kern 0.5pt%
          \raise 1pt
          \vbox{\hrule width3.6pt height0.3pt depth0pt
            \kern -1.5pt
            \hbox{\kern 1.65pt
              \vrule width0.3pt height4.5pt depth0pt
              }
            \kern -1.5pt
            \hrule width3.6pt height0.3pt depth0pt}%
            \kern 0.5pt
        }
    }}

  \def\mm{{\mathchoice
               {
                 \kern 1pt
           \raise 1pt    \vbox{\hrule width5pt height0.4pt depth0pt
                  \kern 2pt
                  \hrule width5pt height0.4pt depth0pt}
                 \kern 1pt}
               {
                \kern 1pt
           \raise 1pt \vbox{\hrule width4.3pt height0.4pt depth0pt
                  \kern 1.8pt
                  \hrule width4.3pt height0.4pt depth0pt}
                 \kern 1pt}
               {
                \kern 0.5pt
           \raise 1pt
                \vbox{\hrule width4.0pt height0.3pt depth0pt
                  \kern 1.9pt
                  \hrule width4.0pt height0.3pt depth0pt}
                \kern 1pt}
               {
               \kern 0.5pt
         \raise 1pt  \vbox{\hrule width3.6pt height0.3pt depth0pt
                  \kern 1.5pt
                  \hrule width3.6pt height0.3pt depth0pt}
               \kern 0.5pt}
               }}

\def\pd{{\kern0.5pt
           + \kern-5.05pt \raise5.8pt\hbox{$\textstyle.$}\kern
0.5pt}}

\def\pmd{{\kern0.5pt
          \pm \kern-5.05pt
\raise6.3pt\hbox{$\textstyle.$}\kern1.5pt}}

\def\md{{\mathchoice
   {
      {{\kern 1pt - \kern-6.2pt \raise5pt\hbox{$\textstyle.$}\kern
1pt}}}
    {
      {{\kern 1pt - \kern-6.2pt \raise5pt\hbox{$\textstyle.$}\kern
1pt}}}
    {
      {\kern0.5pt - \kern-5.05pt
\raise3.4pt\hbox{$\textstyle.$}\kern0.5pt}}
    {
      {\kern0.5pt - \kern-5.05pt
\raise3.4pt\hbox{$\textstyle.$}\kern0.5pt}}}}

\def\beq{\begin{equation}}
\def\eeq{\end{equation}}
\def\bea{\begin{eqnarray}}
\def\eea{\end{eqnarray}}
\def\Tr{\textstyle{Tr}}
\def\tr{\textstyle{tr}}
\def\a{\alpha}
\def\b{\beta}
\def\g{\gamma}

\def\d{\delta}
\def\e{\epsilon}

\def\th{\theta}

\def\G{\Gamma}
\def\D{\Delta}

\def\L{\Lambda}
\def\f{\phi}

\begin{document}

\section{Introduction}

In the last few years, AdS/CFT correspondence and stringy--inspired
technologies for computing scattering amplitudes have led to the
discovery of new remarkable properties of supersymmetric Yang--Mills
theories in four dimensions.

For planar ${\cal N}=4$ SYM theory, a duality between MHV scattering
amplitudes and light--like polygon Wilson loops has been found first
at strong coupling \cite{AM} and successively tested at weak coupling
by a perturbative field theory approach
\cite{Drummond:2007aua}-\cite{Anastasiou:2009kna}. On the field theory
side this duality is related to the emergence of a new hidden
symmetry, the dual superconformal symmetry
\cite{Drummond:2008vq,Brandhuber:2008pf}, which corresponds at strong
coupling to the invariance of the type IIB string theory on ${\rm
  AdS}_5 \times {\rm S}_5$ under a suitable combination of bosonic and
fermionic T--dualities \cite{BM}.  The dual superconformal generators
are part of the infinite set of generators of the Yangian symmetry of
the theory \cite{Drummond:2009fd}, thus being intimately related to
its integrability \cite{integrability}-\cite{Alday:2010vh}.

More recently, a novel duality has been discovered \cite{AEKMS} which
involves correlation functions of gauge invariant BPS scalar operators
of ${\cal N}=4$ SYM theory. An $n$--point correlation function ${\cal
  C}_n$ in the limit where adjacent points become light--like
separated should be equal to a light--like $n$--polygon Wilson loop in
the adjoint representation of the gauge group. The precise
identification
\begin{equation}
  \lim_{x_{i,i+1}^2 \to 0} \, \frac{ {\cal C}_n}{{\cal C}_n^{~tree}} = \langle \Tr_{adj} P \exp{ \left(
      ig \, \int_{\G_n} dz^\mu A_\mu(z) \right)} \rangle
\end{equation}
has been tested perturbatively up to two loops in a number of cases
\cite{AEKMS}.

While this new duality is still lacking a proof in the string theory
regime, in field theory a physical explanation can be given in terms
of an infinitely fast moving scalar particle interacting with a low
energy gluon. In the light--like limit, in fact, the scalar particle
flowing around the loop becomes infinitely energetic compared to the
gluon it interacts with. As a consequence, its propagator becomes an
almost free propagator, except for an eikonal phase $P \exp{ \left(ig
    \, \int_{\G_n} dz^\mu A_\mu(z) \right)}$ which arises as the
result of a path integral saddle point approximation.  According to
this explanation, the connection between correlators and polygonal
Wilson loops should be true not only for ${\cal N}=4$ SYM but also for
general conformal gauge theories in any dimensions \cite{AEKMS}.

Since Wilson loops are dual to planar scattering amplitudes, a direct
duality between $n$--point correlation functions and $n$--point
scattering amplitudes must exist. This has been investigated in
\cite{EKS}, where this duality has been established at one--loop for
generic $n$ and proved at two loops for $n=4,5,6$.

It is interesting to investigate whether the amplitudes/WL/correlators
dualities and the existence of underlying hidden symmetries extend to
class of theories in different dimensions for which a string dual
description is known.

In this paper we concentrate on the class of ${\cal N}=2$, $U(N)_{k_1}
\times U(M)_{k_2}$ Chern--Simons matter theories in three dimensions
with generic $(k_1,k_2)$ CS levels and generic superpotential. It
includes, as particular cases, the ${\cal N}=6$ ABJM theory
\cite{ABJM,ABJ} dual to a IIA string theory on ${\rm AdS}_4 \times
{\rm CP}_3$, the ${\cal N}=8$ BLG theory \cite{BL,G} describing the
low--energy dynamics of M2--branes in M--theory and ${\cal N}=2,3$
superconformal theories \cite{BPS1, BPS2} for which a dual description
in terms of deformed backgrounds has been conjectured \cite{GT}.

For the ABJM theory, preliminary results are already available in the
literature.  At tree level, scattering amplitudes exhibit dual
superconformal symmetry \cite{HL, GHKLL} whose generators are the
level--one generators of a Yangian symmetry.  Invariance under Yangian
symmetry has been explicitly proved at tree level for four and
six--point amplitudes \cite{BLM} and argued in general through the construction 
of a generating function \cite{Lee:2010du}. These symmetries suggest that string
theory on ${\rm AdS}_4 \times {\rm CP}_3$ should be self--dual under a
suitable combination of bosonic and fermionic T--dualities. Efforts in
this direction are complicated by the emergence of singularities
\cite{ADO}--\cite{DO}.

A first indication of a duality between scattering amplitudes and
Wilson loops comes from the fact that both the one--loop four--point
amplitude \cite{ABM} and the light--like square WL \cite{HPW} vanish.

In this paper we move one step further in the direction of
establishing amplitudes/WL/correlators dualities, by studying
correlation functions of gauge invariant BPS scalar operators.

For generic ${\cal N}=2$, two--level Chern--Simons matter theories we
compute the $n$--point correlator at one--loop.  We prove that in the
light--like limit its expression, divided by the corresponding tree
level correlator, coincides with the one--loop expression for a
light--like $n$--polygon Wilson loop, once the Feynman combining
parameters of the correlator integral are identified with the affine
parameters which parametrize the light--like edges of the WL
polygon. Remarkably, we find that both quantities can be expressed in
terms of a five dimensional two--mass--easy box integral.

While in the ABJM case, and whenever $K_2=-K_1$ and $M=N$, the
identification gets trivialized by the fact that both the correlator
and the Wilson loop are proportional to a vanishing color factor, in
the more general cases the color factor in front is not zero and a
non--trivial identification emerges.

We manage to compute the five dimensional box integral analytically
and prove that, once plugged back into the correlation function/WL, it
gives a vanishing result.

Our final statement is then
\begin{equation}
  \lim_{x_{i,i+1}^2 \to 0} \, \frac{ {\cal C}_n^{~1-loop}}{{\cal C}_n^{~tree}} =  
  \langle W_n \rangle^{1 -loop} ~=~ 0 \qquad , \qquad {\rm for ~any~} n
  \label{result}
\end{equation}
This identity is true for any value of the CS levels and for $N,M$
finite (no planar limit is required).  It holds for the whole class of
theories under study, independently of their degree of
supersymmetry. This is a consequence of the fact that at the order we
are working, the superpotential does not enter the calculation. Note
that at one loop they are all superconformal theories, being the
beta--functions trivially zero \cite{BPS1, BPS2}. We expect that
theories with different number of supersymmetries and with or without
superconformal invariance will undergo a different destiny starting
from two--loops \cite{preparation1}.

We stress that in the general case the identification between
correlators and Wilson loops is valid independently of the fact that
they both eventually vanish.  Therefore, our result is a first
non--trivial indication of a correlator/WL duality at work and
supports the conjecture of \cite{AEKMS} which states that this duality
should hold for generic conformal gauge theories in any dimensions.

For ${\cal N} >4$, four--point scattering amplitudes have been proved
to vanish at one loop \cite{ABM}. Therefore, for the special case
$n=4$, our result completes the amplitudes/WL/correlators duality for
theories with a number of supersymmetries greater than four.

We prove that the $n$--polygon Wilson loop is zero at first order for
any value of the CS levels and independently of the chiral
couplings. Thus, the proof is true also for pure Chern--Simons
theories, just set matter fields and one of the two gauge fields to
zero.  Therefore, our result provides the analytical proof of the
conjecture made in \cite{HPW} according to which one--loop light--like
Wilson loops should vanish in pure Chern--Simons theories.

The paper is organized as follows. In Section 2 we introduce the class
of CS matter theories we are interested in and list the corresponding
gauge invariant chiral operators. Working in ${\cal N}=2$ superspace,
we classify different theories according to the particular choice of
the coupling constants in the superpotential. In Section 3 we focus on
the evaluation of $n$--point correlators for dimension--one chiral
operators. In particular, we evaluate the building block which enters
one--loop calculations and discuss its representation in terms of a 5d
two--mass easy box integral.  In Section 4 we prove that in the
light--like limit, the expression for the one--loop correlator divided
by its tree level counterpart is identical to the first order
contribution to a light--like $n$--polygon Wilson loop. This
identification holds independently of the value of the couplings and
even before computing the actual Feynman integrals. In Section
5, equipped with the exact result for the 5d box integral, we give the
analytical proof that correlators and Wilson loops vanish at this
order. In Section 6 we prove that our results for dimension--one BPS
operators extend to correlation functions of operators with arbitrary
dimension.  Conclusions with a discussion of future perspectives
follow, plus Appendix A which contains our conventions and Appendix B
where we present a detailed discussion of the unexpected emergence of
a 5d box integral.

\section{${\cal N}=2$ Chern--Simons matter theories}

In three dimensions, we consider a ${\cal N}=2$ supersymmetric $U(N)
\times U(M)$ Chern--Simons theory for vector multiplets $(V,\hat{V})$
coupled to chiral multiplets $A^i$ and $B_i$, $i=1,2$ in the
fundamental representation of a global $SU(2)_A \times SU(2)_B$. The
vector multiplets $V, \hat{V}$ are in the adjoint representation of
the gauge groups $U(N)$ and $U(M)$ respectively, while $A^i$ are in
the $(N,\bar{M})$ and $B_i$ in the $(\bar{N},M)$ bifundamental
representations.

In ${\cal N}=2$ superspace the action reads (for superspace
conventions see Appendix A)
\begin{equation} {\cal S} = {\cal S}_{\mathrm{CS}} + {\cal
    S}_{\mathrm{mat}}
  \label{eqn:action}
\end{equation}
with
\begin{eqnarray}
  \label{action}
  && {\cal S}_{\mathrm{CS}}
  =  \int d^3x\,d^4\theta \int_0^1 dt\: \Big\{ K_1 \Tr \Big[
  V \Db^\a \left( e^{-t V} D_\a e^{t V} \right) \Big]
  + K_2  \Tr \Big[ \hat{V} \Db^\a \left( e^{-t \hat{V}} D_\a
    e^{t \hat{V}} \right) \Big]   \Big\}
  \non \\
  \non \\
  && {\cal S}_{\mathrm{mat}} = \int d^3x\,d^4\theta\: \Tr \left( \bar{A}_i
    e^V A^i e^{- \hat{V}} + \bar{B}^i e^{\hat V} B_i
    e^{-V} \right)
  \non \\
  &~& ~ +\int d^3x\,d^2\theta\:
  \Tr  \left[ h_1 (A^1 B_1)^2 + h_2 (A^2 B_2)^2 + h_3 (A^1 B_1 A^2 B_2)
    + h_4 (A^2 B_1 A^1 B_2) \right]  + \, h.c.
  \non\\
\end{eqnarray}
Here $4\pi K_1$, $4\pi K_2$ are two independent integers, as required
by gauge invariance of the effective action.  In the perturbative
regime we take $K_1, K_2 \gg N,M$.

For generic values of the couplings, the action (\ref{eqn:action}) is
invariant under the following gauge transformations
\begin{eqnarray}
  && e^V \rightarrow e^{i \bar{\L}_1} e^V e^{-i\L_1}
  \qquad \qquad e^{\hat{V}} \rightarrow e^{i\bar{\L}_2} e^{\hat{V}} e^{-i\L_2}
  \\
  && \non \\
  && A^i \rightarrow e^{i\L_1} A^i e^{-i\L_2}
  \qquad \qquad B_i \rightarrow e^{i\L_2} B_i e^{-i\L_1}
  \label{gaugetransf}
\end{eqnarray}
where $\L_1, \L_2$ are two chiral superfields parametrizing $U(N)$ and
$U(M)$ gauge transformations, respectively. Antichiral superfields
transform according to the conjugate of (\ref{gaugetransf}).  The
action is also invariant under the $U(1)_R$ R--symmetry group under
which the $A^i$ and $B_i$ fields have charges $\frac12$.

For special values of the couplings we can have enhancement of global
symmetries and/or R--symmetry with consequent enhancement of
supersymmetry.

For $K_1 = - K_2$ and $h_1 = h_2 = 0$ we have ${\cal N}=2$
ABJM/ABJ--like CFT's \cite{BPS1}.  In this case the theory is
invariant under two global $U(1)$'s
\begin{eqnarray}
  && U(1)_A: \quad A^1 \rightarrow e^{i\a} A^1  \qquad ~, \qquad  U(1)_B: \quad B_1 \rightarrow e^{i\b} B_1
  \non \\
  && ~ \qquad \qquad A^2 \rightarrow e^{-i\a} A^2 \qquad , \qquad \qquad  \qquad ~ B_2 \rightarrow e^{-i\b} B_2
  \label{U(1)}
\end{eqnarray}
If, in addition, we choose $h_3 = - h_4 \equiv h$, the global symmetry
becomes $U(1)_R \times SU(2)_A \times SU(2)_B$ and gets enhanced to
$SU(4)_R$ for $h = 1/K$ \cite{ABJM, klebanov}. For this particular
values of the couplings we recover the ${\cal N}=6$ superconformal ABJ
theory \cite{ABJ} and for $N=M$ the ABJM theory \cite{ABJM}.

In the more general case $K_1 \neq -K_2$, setting $h_1 = h_2 = \frac12
\left( h_3 + h_4 \right)$ the corresponding superpotential reads
\begin{equation}
  {\cal S}_{\mathrm{pot}} = \frac12 \, \int
  d^3x\,d^2\theta\: \Tr \left[ h_3 (A^i B_i)^2 + h_4 (B_i A^i)^2
  \right] \, + \, h.c.
  \label{potGT}
\end{equation}
This is the class of ${\cal N}=2$ theories studied in \cite{GT} with
$SU(2)$ invariant superpotential, where $SU(2)$ rotates simultaneously
$A^i$ and $B_i$.

When $h_3 = -h_4$, that is $h_1 = h_2 = 0$, we have the particular set
of ${\cal N}=2$ theories with global $SU(2)_A \times SU(2)_B$ symmetry
\cite{GT}. This is the generalization of ABJ/ABJM--like theories to
$K_1 \neq -K_2$.  For particular values of the couplings \cite{BPS2}
the theory is superconformal invariant.

Finally, another interesting fixed point corresponds to $h_3 =
\frac{1}{K_1}$ and $h_4 = \frac{1}{K_2}$.  The $U(1)_R$ R--symmetry is
enhanced to $SU(2)_R$ and the theory is ${\cal N}=3$ superconformal
\cite{GT,BPS2}.

The quantization of the theory can be easily carried on in superspace
after performing gauge fixing (for details, see for instance
\cite{BPS1, BPS2}). In coordinate space and using Landau gauge, this
leads to gauge propagators
\begin{eqnarray}
  \langle V^A(1) \, V^B(2) \rangle
  =   \frac{1}{4\pi K_1} \, \Db^\a D_\a \, \frac{\delta^4(\th_1-\th_2)}{|x_1 - x_2|} \, \delta^{AB} \nonumber \\ \langle \hat
  V^A(1) \, \hat V^B(2) \rangle =
  \frac{1}{4\pi K_2} \, \Db^\a D_\a  \, \frac{\delta^4(\th_1-\th_2)}{|x_1 - x_2|} \, \delta^{AB}
  \label{gaugeprop}
\end{eqnarray}
whereas the scalar propagators are
\begin{eqnarray}
  &&\langle \bar A^{\hat a}_{\ a}(1) \, A^b_{\ \hat b}(2) \rangle
  = \frac{1}{4\pi}  \, D^2 \bar{D}^2 \, \frac{\delta^4(\th_1 - \th_2)}{|x_1 - x_2|} \, \delta^{\hat a}_{\ \hat
    b} \, \delta^{\ b}_{a}
 \nonumber \\
  &&  \langle \bar B^a_{\ \hat a}(1) \, B^{\hat b}_{\ b}(2) \rangle =
  \frac{1}{4\pi} \, D^2 \bar{D}^2 \, \frac{\delta^4(\th_1 - \th_2)}{|x_1 - x_2|}  \, \delta^a_{\ b} \, \delta^{\ \hat b}_{\hat a}
  \label{scalarprop}
\end{eqnarray}
The vertices needed for one--loop calculations can be easily read from
the action (\ref{action})
\begin{equation}
  \label{vertices}
  \int d^3x\,d^4\theta\: \left[ \Tr ( \bar{A}_i V A^i) -  \Tr ( B_i V \bar{B}^i )
    + \Tr (  \bar{B}^i {\hat V} B_i ) -  \Tr ( A^i {\hat{V}}   \bar{A}_i ) \right]
\end{equation}
We note that $A$ and $B$ vertices always carry an opposite sign.

\vskip 20pt
In ${\cal N}=2$ superspace language, the most general gauge invariant,
BPS scalar operator is
\begin{equation}
  {\cal O}^{i_1, \cdots , i_l}_{j_1, \cdots , j_l} =
  \Tr (A^{i_1} B_{j_1} \cdots A^{i_l} B_{j_l} )
  \label{BPS}
\end{equation}
It has classical dimension $\D = l$ and belongs to a suitable
representation of $SU(2)_A \times SU(2)_B$.  Indeed, according to the
particular theory we are considering, the sequence of indices may be
constrained by the request for the operator to be a chiral primary
(${\cal O}^{i_1, \cdots , i_l}_{j_1, \cdots , j_l} \neq \bar{D}^2
{\cal X}^{i_1, \cdots , i_l}_{j_1, \cdots , j_l} $). In the ABJ/ABJM
case, this amounts to require the indices to be completely
symmetrized, as follows from the observation that the equations of
motion set antisymmetric products equal to $\bar{D}^2$(something).

For theories with $U(1)_A \times U(1)_B$ invariance (\ref{U(1)}), the
composite operator ${\cal O}^{i_1, \cdots , i_l}_{j_1, \cdots , j_l}$
is not in general invariant, unless it contains the same number of
$A^1$ and $A^2$ and the same number of $B^1$ and $B^2$ as well.

 \section{The $n$--point correlation functions}

 We are interested in computing correlation functions of the scalar
 composite operators in (\ref{BPS}). We begin by considering the
 simplest case of a dimension--one operator
\begin{equation}
  {\cal O}^i_j (Z) =  \Tr (A^i (Z) B_j(Z) ) \qquad , \qquad \bar{{\cal O}}_i^j(Z) =  \Tr (\bar{A}_i (Z) \bar{B}^j(Z) )
  \label{BPS1}
\end{equation}
Here $Z = (x^\mu, \th^\a, \thb^{\adot})$ and $i,j$ are flavor indices
that we omit in what follows. The generalization to 
higher dimensional operators is discussed in Section 6, where we prove that 
one--loop correlation functions for BPS operators of arbitrary dimension
can be expressed in terms of  one--loop
correlation functions of dimension--one operators.

We focus on the evaluation of the expression
\begin{eqnarray}
  {\cal C}_{n} = \left. \left\langle   {\cal O}(Z_1)\, \bar{{\cal O}}(Z_2) \cdots
      {\cal O}(Z_{n-1})\, \bar{{\cal O}}(Z_{n})   \right\rangle
  \right|_{\th_i = \thb_i = 0}
  \label{corr}
\end{eqnarray}
which corresponds to the correlator for the lowest scalar component of
(\ref{BPS1}).

At tree level, the correlation function is given by the product of
free chiral propagators (\ref{scalarprop}) which, evaluated at $\th =
\thb = 0$, are simply $\frac{1}{4 \p}\frac{1}{|x_i-x_j|}$. Taking into
account all the possibilities of contracting the fields, the expression
(\ref{corr}) will be a linear combination of connected and
disconnected diagrams. We concentrate only on the connected
part. Using the simplified notation $x_{i,\,j}= |x_i-x_j|$, the tree
level connected correlator reads
\begin{eqnarray}
  {\cal C}_{n}^{~tree} = \frac{MN}{(4 \p)^n} \sum_{\{i_1, \cdots , i_n\}} \frac{1}{x_{i_1,\,i_2}} \frac{1}{x_{i_2,\,i_3}}
  \cdots \frac{1}{x_{i_n,\,i_1}}
  \label{connected}
\end{eqnarray}
where the sum is over all non--cyclic permutations compatible with the
constraint that contractions are allowed only between chirals and
antichirals. Since we will be eventually interested in the behavior of
the correlator in the light--cone limit $x_{i,\,i+1}^2 \to 0$, in
(\ref{connected}) we select the most singular term which corresponds
to the cyclic order $\{1,2, \cdots, n\}$
\begin{eqnarray}
  {\cal C}_{n}^{~tree} \rightarrow \frac{MN}{(4 \p)^n} \prod_{i=1}^{n} \frac{1}{x_{i,\,i+1}}
  \label{cyclic}
\end{eqnarray}
where $x_{n+1}=x_1$.\\
Diagrammatically, this is given by a planar $n$--polygon with the
operators at the vertices (See Fig. \ref{short-correlator}).
\FIGURE{
  \centering
  \includegraphics[width = 0.35 \textwidth]{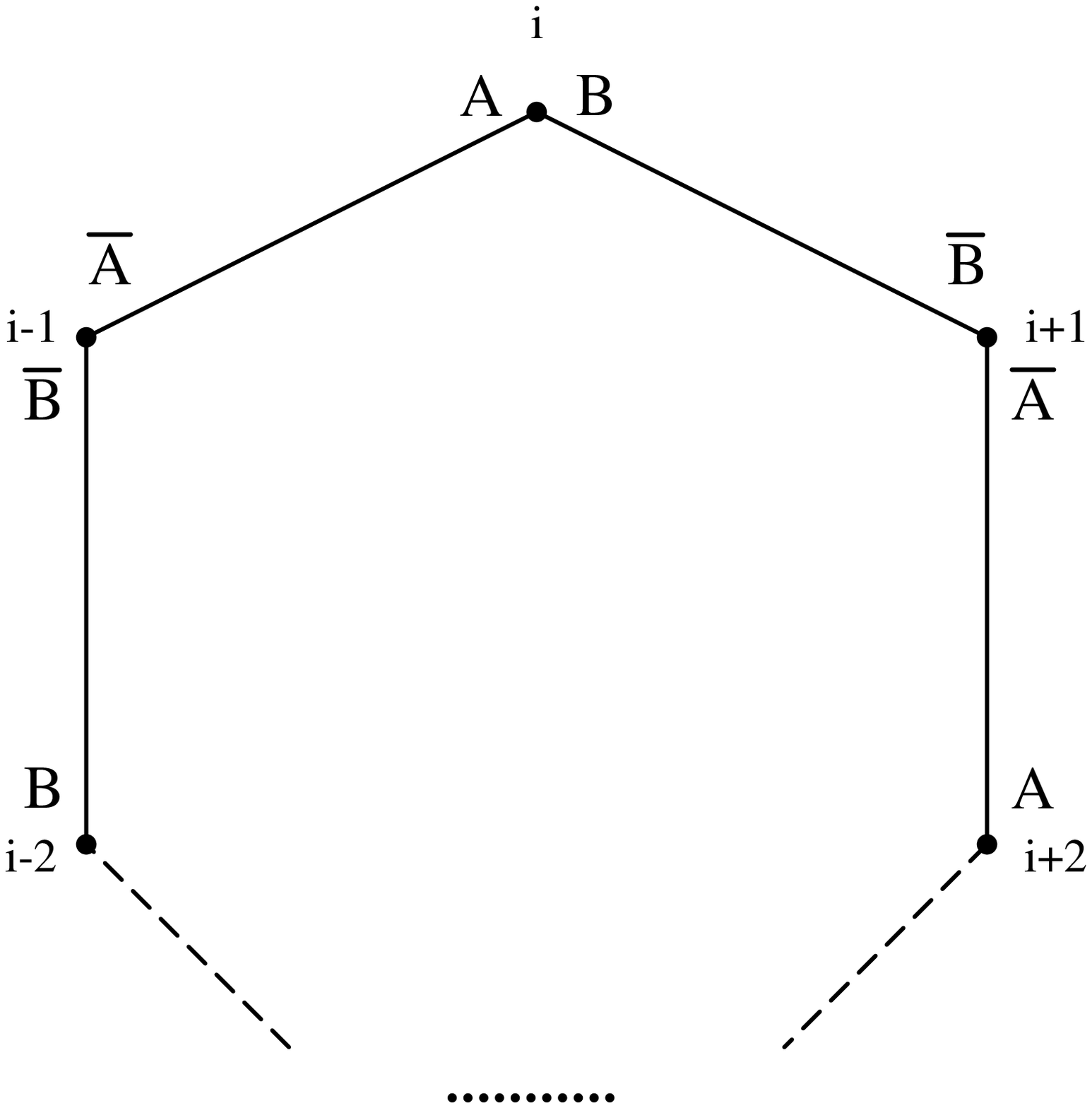}
  \caption{The correlation function of dimension--one operators in the
    light--like limit.}
  \label{short-correlator}
}

First order corrections in the $\frac{N}{K_1}$, $\frac{M}{K_2}$
couplings are obtained by attaching a $V$ or $\hat{V}$ gauge
propagator at the edges of the polygon in all possible ways. At this
order, chiral interaction vertices from the superpotential do not
contribute, so the results are valid for any ${\cal N}=2$ theory.

When the gauge propagator has both ends attached to a single chiral
line the result is zero. In fact, one loop corrections to chiral
propagators vanish because it is possible to perform D--algebra in
such a way that no enough spinorial derivatives survive inside the loop.  

We are then left with contributions where the gauge propagator joins
two different edges.  It is useful to compute the generic building
blocks ${\cal B}_{ij}$ depicted in Fig. \ref{fig:buildingblock}, where
the edges $x_{i,\,i+1}$ and $x_{j,\,j+1}$ are connected by a wave line
representing either a $V$ or a $\hat{V}$ propagator.
\FIGURE{
  \centering
  \includegraphics[width = 0.85 \textwidth]{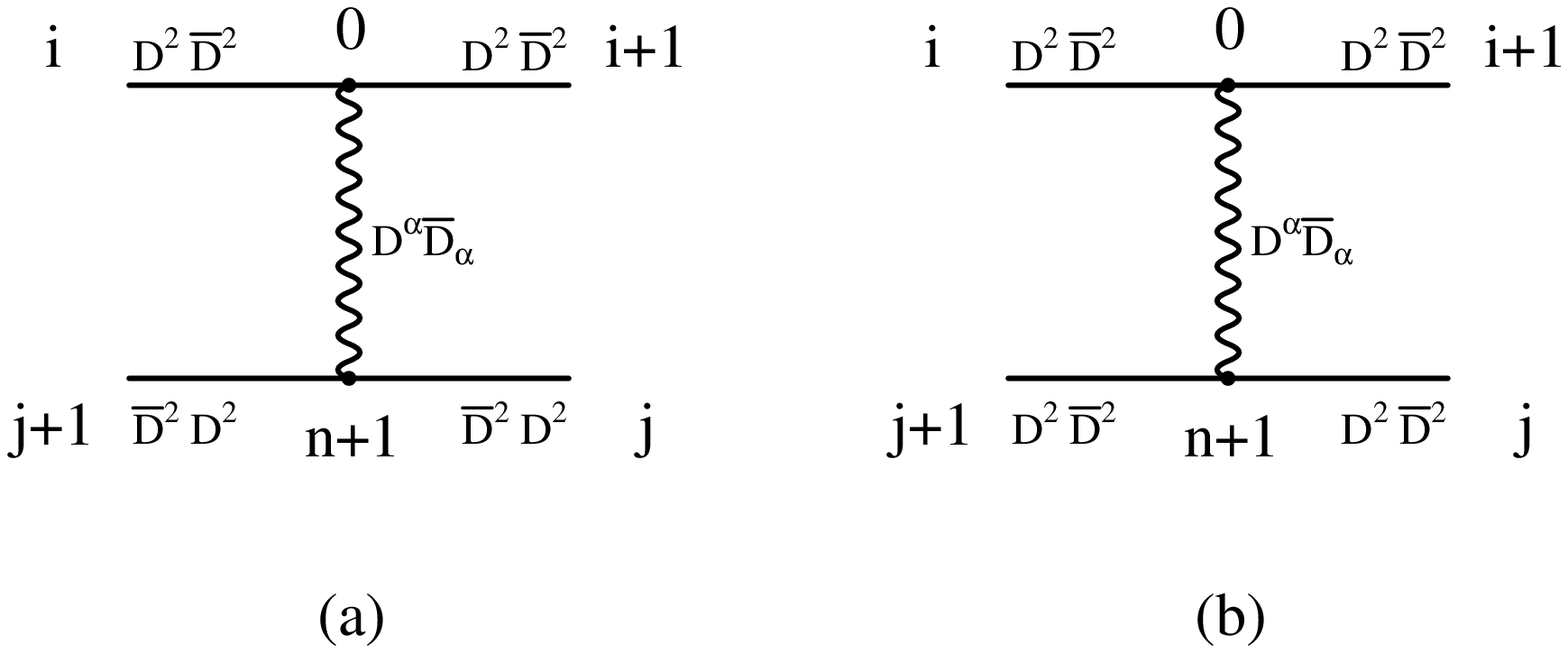}
  \caption{Building blocks for one--loop corrections.}
  \label{fig:buildingblock}
}

\subsection{One--loop building block }

As shown in Fig. \ref{fig:buildingblock}, there are two different
configurations for the one--loop building block, depending on the
chirality of the external vertices. Diagram \ref{fig:buildingblock}a)
corresponds to the case where vertices $i$ and $j$ are antichirals and
$i+1$ and $j+1$ chirals. Diagram \ref{fig:buildingblock}b)
corresponds to the case where vertices $i$ and $j+1$ are antichirals,
while the other two are chirals.

In order to evaluate the building blocks ${\cal B}_{ij}$ we need
perform $D$--algebra to end up with a non--vanishing result when
evaluated at $\th_k=\thb_k=0$, $k=i,i+1,j,j+1 $. Starting with the
configurations of Fig. \ref{fig:buildingblock} for the spinorial
derivatives, we free the gauge and one of the chiral lines from
derivatives by integrating by parts at one of the vertices. Among
different terms which get produced, the only non--trivial contribution
in the $\th_k=\thb_k=0$ limit is the one where a $D^2 \bar{D}^2$
structure survives on three chiral propagators. Together with the
derivatives coming out from the spinorial integrations, these
derivatives are sufficient to kill the fermionic delta functions,
leading to a non--vanishing result.  As a result of the $D$--algebra,
the ordinary Feynman diagram we are left with has three space--time
derivatives acting on chiral propagators.

Summing the contributions from the $V$ and $\hat{V}$ insertions, the
final result for the two configurations is
\begin{eqnarray}
  \label{block}
  && {\cal B}_{ij}^{(a)} = - \frac{2}{(4 \p)^5}   \,
  \left( \frac{1}{K_1} + \frac{1}{K_2} \right) \, \e_{\m\n\rho} \, \partial_{i}^{\m} \, \partial_{i+1}^{\n} \, \partial_{j+1}^{\rho}\, {\cal I}(i,j)
  \non \\
  && {\cal B}_{ij}^{(b)} = \, \, \frac{2}{(4 \p)^5}   \,
  \left( \frac{1}{K_1} + \frac{1}{K_2} \right) \, \e_{\m\n\rho}\, \partial_{i}^{\m} \, \partial_{i+1}^{\n} \, \partial_{j+1}^{\rho}\, {\cal I}(i,j)
\end{eqnarray}
in terms of the integral
\begin{eqnarray}
  \label{integral}
  {\cal I}(i,j) = \int \, \frac{d^3
    x_0\, d^3 x_{n+1}}{x_{0,\,i}\, x_{0,\,i+1} \, x_{0,\, n+1}
    \,x_{j,\,n+1}\, x_{j+1,\,n+1}}
\end{eqnarray}
The remarkable fact is that the expression
$\e_{\m\n\rho}\, \partial_{i}^{\m}\, \partial_{i+1}^{\n}\, \partial_{j+1}^{\rho}\,
{\cal I}(i,j)$ can be manipulated by using Feynman combining and
Mellin--Barnes representation and reduced to a single integral in five
dimensions. Precisely, as proved in details in Appendix \ref{AppB},
the following identity holds
\begin{equation}
  \e_{\m\n\rho}\, \partial_{i}^{\m}\, \partial_{i+1}^{\n}\, \partial_{j+1}^{\rho}\, {\cal I}(i,j) = \frac{8}{\pi^2}\,
  \frac{\e_{\m\n\rho}\, x_{i,\,i+1}^{\m}\, x_{i+1,\,j}^{\n} \, x_{j,\,j+1}^{\rho}}{ x_{i,\,i+1}\, x_{j,\,j+1}}
  \times   \int d^5 x_0\, \frac{1}{x_{0,i}^2\, x_{0,i+1}^2\, x_{0,j}^2\, x_{0,j+1}^2}
  \label{5dint}
\end{equation}
Therefore, the building block which describes the gauge correction to
the tree level expression $\frac{1}{x_{i,\,i+1}\, x_{j,\,j+1}}$ can be
still written as the product of the two free propagators times a five
dimensional scalar integral.  Interpreting the $x_j$ variables as the
dual coordinates of a set of 5d momenta $p_j = x_{j+1}- x_j$, this can
be seen as a box integral in five dimensions.

\subsection{One--loop results and their light--like limit}
\label{sec:oneloop}

Given the results (\ref{block}, \ref{5dint}), we are now ready to
evaluate the one--loop $n$--point correlator.  The generic
contribution will be the product of the blocks (\ref{block}) times
$(n-2)$ free propagators.

By antisymmetry of the $\e$ tensor we can ascertain that contributions
coming from the gauge propagator connecting two adjacent edges vanish
identically. In fact, setting $x_i = x_{j+1}$ or $x_j = x_{i+1}$, it
is immediate to see that the structure $\e_{\m\n\rho}\,
x_{i,\,i+1}^{\m}\, x_{i+1,\,j}^{\n} \, x_{j,\,j+1}^{\rho}$ is
zero. Therefore, we are left only with contributions where the gauge
propagator connects two non--adjacent edges.

When the two lines are separated by an odd number of free propagators
the block ${\cal B}^{(a)}_{ij}$ has to be used.  In this case, given
the particular structure of the operator and the fact that only the
$\langle A \bar{A} \rangle$, $\langle B \bar{B} \rangle$ propagators
are non--vanishing, the vertices employed to construct the block are necessarily of the same
type: If one is a $A$--vertex, the second one is a $A$--vertex as
well. These carry the same sign so that this contribution is given by
${\cal B}^{(a)}_{ij}$, without any sign change.  On the other hand,
when the two non--adjacent lines are separated by an even number of
free lines we need use the block ${\cal B}^{(b)}_{ij}$. In this case
the two employed vertices are of different kind and since these carry
opposite sign (see eq. (\ref{vertices})) we obtain an extra minus
which compensates the sign difference between the blocks, so that both kinds of 
contributions end up having the same sign. 

In conclusion, taking into account color factors, the leading
term of the correlation function at one--loop is
\begin{eqnarray}
  \label{sum}
  {\cal C}_{n}^{~1-loop} \rightarrow  {\cal C}_n^{tree}\,  \times  \frac{-1}{ 4\, \p^5}\, \left[ \frac{N}{K_1} + \frac{M}{K_2} \right]\, \sum_{i=1}^{n-2}\,\, \sum_{j=i+2}^{n-\d_{i,1}} \,\e_{\m\n\rho}\, x_{i,\,i+1}^{\m}\, x_{i+1,\,j}^{\n}\, x_{j,\,j+1}^{\rho}\, {\cal J}(i,j)
\end{eqnarray}
where the sum extends to the $n(n-3)/2$ ways to connect two
non--adjacent edges, and ${\cal J}(i,j)$ is
\begin{equation}
  \label{5easy}
  {\cal J}(i,j) =   \int d^5 x_0\, \frac{1}{x_{0,i}^2 \, x_{0,i+1}^2 \, x_{0,j}^2 \, x_{0,j+1}^2}
\end{equation}

In the ABJM case and for all theories with $K_2 = - K_1$ and $M=N$ the
color factor in front vanishes, so that correlation functions are
trivially zero at one loop.  The same happens for the BLG theory, as
it can be easily checked by computing the color factor for gauge group
$SU(N) \times SU(M)$ which turns out to be $ ( N - 1/N ) / K_1 + ( M -
1/M ) / K_2$.

We concentrate on more general theories for which the color factor
does not vanish.  The first non--trivial expression in (\ref{sum}) is the
four point correlation function. Setting $n=4$, the sum reduces to two contributions having the same
integral factor
\begin{eqnarray}
  {\cal C}_{4}^{~1-loop} \propto \e_{\m\n\rho}
  \left(x_{1,\,2}^{\m}\, x_{2,\,3}^{\n}\, x_{3,\,4}^{\rho} +
    x_{2,\,3}^{\m}\, x_{3,\,4}^{\n}\, x_{4,\,1}^{\rho}\right)\, \int
  d^5 x_0\, \frac{1}{x_{0,1}^2 \, x_{0,2}^2 \, x_{0,3}^2 \, x_{0,4}^2}
\end{eqnarray}
It is immediate to see that the structure in front of the integral
vanishes, due to the contraction with the $\e$ tensor. Hence, the
connected four point correlation function is identically zero, no need
to perform the integral.

This trivial result is no longer true for higher point correlation
functions, so that in general we really have to work out the ${\cal
  J}(i,j)$ integral.  We do it in the light--like limit $x_{i,\,i+1}^2
\to 0$, which greatly simplifies the computation and, as shown in
\cite{AEKMS}, is the correct prescription to test a correspondence to
light--like Wilson loops.

Since the prefactor ${\cal C}_n^{~tree}$ in (\ref{sum}) is divergent
in this limit, we consider the ratio of the one-loop correlator to the
tree level result.  Moreover, in order to get a real output, we
require the $n(n-3)/2$ diagonals of the $n$--polygon to be
space--like ($x_{i,j}^2 > 0,\, j \neq i+1$).

To evaluate the integral (\ref{5easy}) we first shift the integration
variable $x_0 \rightarrow x_0 + x_i$, and reduce it to a Feynman
scalar box integral in five dimensions with external momenta
$x_{i,\,i+1}$, $x_{i+1,\,j}$, $x_{j,\,j+1}$ and $x_{j+1,\,i}$. In the
light--like limit the integral is recognized to be the two mass easy
box, with two of the momenta massless by construction and the other
two massive.  In the $j=i+2$ case, i.e. when the two edges are
separated by a single free line, one more external leg becomes
massless and the integral simplifies further.

Feynman parametrizing the scalar five dimensional box and performing
the loop integration yields
\begin{equation}
  {\cal J}(i,j) = \frac{\pi^3}{2} \int_0^1 [d\a]_4
  \frac{1}{\left(
        \a_1\, \a_3\, x_{i,j}^2 
      + \a_2\, \a_4\, x_{i+1,j+1}^2
      + \a_1\, \a_4\, x_{i,j+1}^2 
      + \a_2\, \a_3\, x_{i+1,j}^2
    \right)^{\frac32}}
\end{equation}
where $[d\a]_4 = \d(1-\sum_{k=1}^4 \a_k) \prod_{k=1}^4 d\a_k$.

The delta--function constraint can be solved by performing the
following change of variables
\begin{equation}
  \a_1 = (1-\b_1)(1-\b_3) 
  \quad , \quad 
  \a_2 =    \b_1 (1-\b_3)
  \quad , \quad
  \a_3 = (1-\b_2)   \b_3 
  \quad , \quad 
  \a_4 =    \b_2    \b_3
\end{equation}
Consequently, the integral reduces to
\begin{eqnarray}
  &&{\cal J}(i,j) =  \frac{\pi^3}{2}\, \int_0^1 \prod_{i=1}^3 d\b_i \, \times
  \\
  && \frac{\b_3^{-\frac12} (1-\b_3)^{-\frac12}}{\left[(1-\b_2)\, (1-\b_1)\, x_{i,j}^2 + \b_1\, \b_2\, x_{i+1,j+1}^2 + \b_2\, (1-\b_1)\, x_{i,j+1}^2 + \b_1\, (1-\b_2)\,
      x_{i+1,j}^2 \right]^{\frac32}}
  \non \\
  \non
\end{eqnarray}
where the $\b_3$--integration can be trivially performed, leading to
\begin{eqnarray}
  \label{5dint3}
  &&{\cal J}(i,j) = \frac{\pi^4}{2} \int_0^1  \, d\b_1 d\b_2 \, \times
  \\
  &&\frac{1}{\left[(1-\b_2)\, (1-\b_1)\, x_{i,j}^2 + \b_1\, \b_2\, x_{i+1,j+1}^2 + \b_2\, (1-\b_1)\, x_{i,j+1}^2 + \b_1\, (1-\b_2)\,
      x_{i+1,j}^2 \right]^{\frac32}}
  \non \\
  \non
\end{eqnarray}
Finally, the last two integrations can be performed with the help of
{\em Mathematica}.

In conclusion, the general one--loop contribution to the $n$--point
correlator corresponding to a Feynman diagram where a vector line
connects the $x_{i,i+1}$ and $x_{j,j+1}$ free propagators, in the
light--cone limit reads
\begin{eqnarray}
  \label{finalresult}
  && \e_{\m\n\rho}\, x_{i,\,i+1}^{\m}\, x_{i+1,\,j}^{\n}\, x_{j,\,j+1}^{\rho}\, {\cal J}(i,j) ~=~ 
  \\
	&&\non\\
   &~& ~~~~
  \pi^4\,  \mathcal{S}_{i,\,j}\,
  \log \left[ \frac{\left( 1 + x_{i+1,j}\, {\cal L}_{i,\,j} \right) \left( 1 + x_{i,j+1}\,
        {\cal L}_{i,\,j} \right)}{\left( 1 - x_{i+1,j}\, {\cal L}_{i,\,j} \right) \left( 1 - x_{i,j+1}\,
        {\cal L}_{i,\,j} \right)}
    \frac{\left( 1 -x_{i,j}\, {\cal L}_{i,\,j} \right) \left( 1 - x_{i+1,j+1}\, {\cal L}_{i,\,j} \right)}{\left( 1 + x_{i,j}\, {\cal L}_{i,\,j} \right) \left( 1 + x_{i+1,j+1}\, {\cal L}_{i,\,j} \right)} \non
  \right] \\ \non
\end{eqnarray}
where we have defined
\begin{equation}
  \label{signs_definition}
  \mathcal{S}_{i,\,j}=
  \frac{2\, \e_{\m\n\rho}\, x_{i,\,i+1}^{\m}\, x_{i+1,\,j}^{\n}\, x_{j,\,j+1}^{\rho}}{\sqrt{x_{i,j}^2 + x_{i+1,j+1}^2 -
      x_{i+1,j}^2 - x_{i,j+1}^2 } \sqrt{x_{i,j}^2 x_{i+1,j+1}^2 - x_{i+1,j}^2  x_{i,j+1}^2 } }
\end{equation}
and
\begin{equation}
  {\cal L}_{i,\,j} = \, \frac{\sqrt{x_{i,j}^2 +
      x_{i+1,j+1}^2 - x_{i+1,j}^2 - x_{i,j+1}^2 }}{\sqrt{x_{i,j}^2
      x_{i+1,j+1}^2 - x_{i+1,j}^2 x_{i,j+1}^2 }}
\end{equation}
Focusing on the argument of the logarithm in (\ref{finalresult}) we
note that it depends only on the diagonals connecting the four
vertices of the block $x_i$, $x_{i+1}$, $x_j$ and $x_{j+1}$, 
as depicted in Fig. \ref{fig:blocdiagonals}(a). 
This is due to the fact that the correlator, being Poincar\'e
invariant, has to be a function of the only invariants that we can construct. 
In the light--like limit these are the
$n(n-3)/2$ space--like diagonals \footnote{Actually not all diagonals are independent and their
  number could in principle be reduced by the Gram constraints. Since
  these constraints are difficult to implement we will not pursue this
  technique.}.  
  
 We distinguish two sets of diagonals. We call ``short"
diagonals those connecting two vertices separated by a pair of
light--like edges, whereas we call ``long" diagonals the remaining $n(n-5)/2$ ones.

An example of the appearance of short diagonals is depicted in
Fig. \ref{fig:blocdiagonals}(b), where the vertices $x_{i+1}$ and $x_j$ are connected 
by a null edge, so the space--like segments $x_{i,\,j}$ and
$x_{i+1,\,j+1}$ are short diagonals. In this case, the corresponding contribution
can be obtained from the general expression
(\ref{finalresult}) by collapsing $x_{i+1,\,j} \to 0$, and as a
result the logarithm contains just three factors instead of four.

\FIGURE{
  \centering
  \includegraphics[width = 0.9 \textwidth]{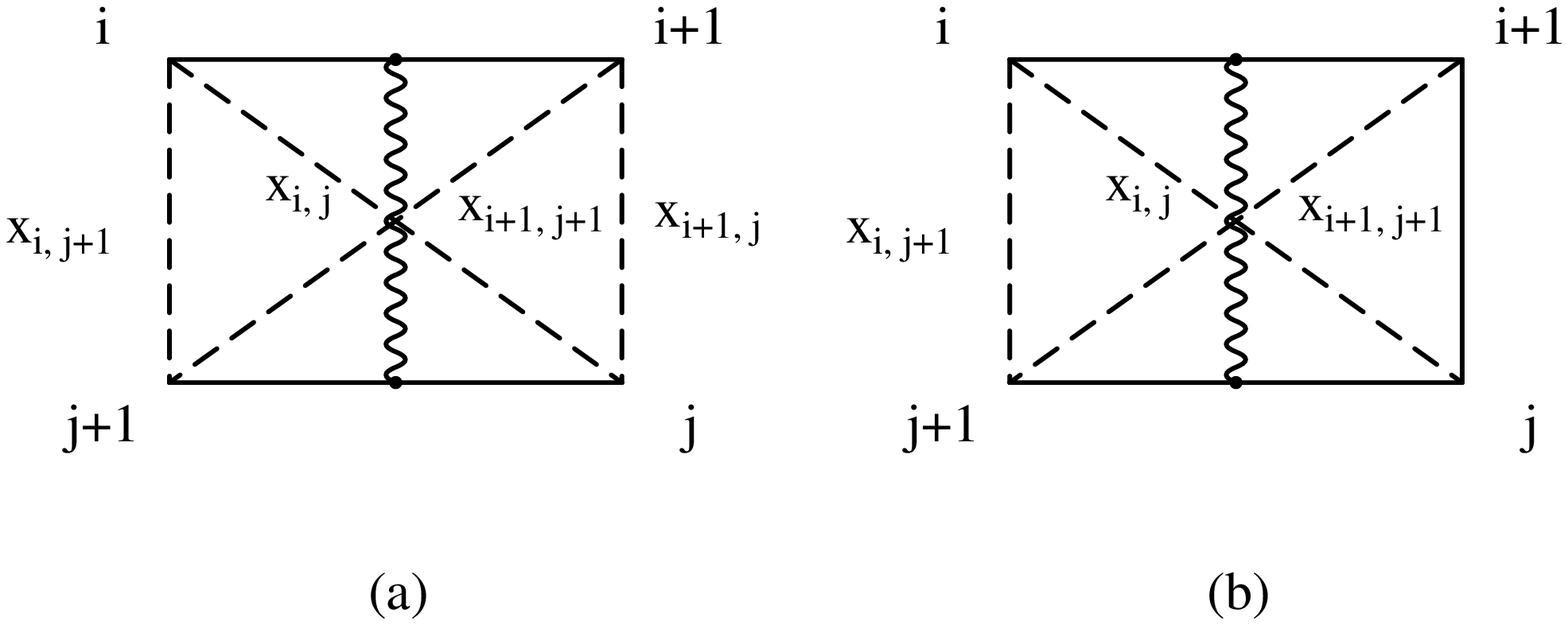}
  \caption{The building blocks for the correlation functions only
    depend on the diagonals of the polygon, which are drawn with
    dashed lines. Case (a) corresponds to the $n(n-5)/2$ blocks where
    all the involved diagonals are long. Case (b) depicts one of the
    $n$ blocks with short diagonals.}
  \label{fig:blocdiagonals}
}

Going back to (\ref{finalresult}), by straightforward algebra
we can rewrite the argument of the logarithm as
\begin{align}
  \label{positiveness}
  &\frac{\left( 1 + x_{i+1,j}\, {\cal L}_{i,\,j} \right) \left(
      1 + x_{i,j+1}\, {\cal L}_{i,\,j} \right)}{\left( 1 -
      x_{i+1,j}\, {\cal L}_{i,\,j} \right) \left( 1 -
      x_{i,j+1}\, {\cal L}_{i,\,j} \right)}
  \frac{\left( 1 -x_{i,j}\, {\cal L}_{i,\,j} \right) \left( 1 - x_{i+1,j+1}\, {\cal L}_{i,\,j} \right)}{\left( 1 + x_{i,j}\, {\cal L}_{i,\,j} \right) \left( 1 + x_{i+1,j+1}\, {\cal L}_{i,\,j} \right)}=\nonumber\\
  &\ \ \ \ \ \ \ \ \ \ \ \ \ \ \ \ \ \ \ \ \ \ \ \ \, \frac{\left( 1 +
      x_{i+1,j}\, {\cal L}_{i,\,j} \right)^2 \left( 1 +
      x_{i,j+1}\, {\cal L}_{i,\,j} \right)^2}{\left( 1 +
      x_{i,j}\, {\cal L}_{i,\,j} \right)^2 \left( 1 +
      x_{i+1,j+1}\, {\cal L}_{i,\,j} \right)^2}
\end{align}
As proven in Section \ref{vanishing},  ${\cal L}_{i,\,j}$'s
are real functions as long as all the diagonals are space--like. Under
this assumption, eq. (\ref{positiveness}) is the square of a
real expression and the logarithm in (\ref{finalresult}) is well defined. A similar
argument applies also to the case of short diagonals,  leading to the
same conclusions. \\

Finally, inserting the result (\ref{finalresult}) back into
eq. (\ref{sum}) and summing over all possible contractions, we obtain
the complete analytical result for the ratio ${\cal C}_n^{~1-loop} /
{\cal C}_n^{tree}$ in the light--like limit. The positiveness of the
arguments of all logarithms allows us to safely rewrite the sum as
\bea\label{finalresult2}
&& \frac{{\cal C}_n^{~1-loop}}{{\cal C}_n^{tree}} = -\frac{1}{ 4\, \p}\, \left[ \frac{N}{K_1} + \frac{M}{K_2} \right] \,
\log \Bigg\{ 
\\ && ~~ \left.
 \prod_{i=1}^{n-2}\,\, \prod_{j=i+2}^{n-\d_{i,1}}\,\, 
\left[ \frac{\left( 1 + x_{i+1,j}\, {\cal L}_{i,\,j} \right) \left( 1 + x_{i,j+1}\,
        {\cal L}_{i,\,j} \right)}{\left( 1 - x_{i+1,j}\, {\cal L}_{i,\,j} \right) \left( 1 - x_{i,j+1}^2\,
        {\cal L}_{i,\,j} \right)}
    \frac{\left( 1 -x_{i,j}\, {\cal L}_{i,\,j} \right) \left( 1 - x_{i+1,j+1}\, {\cal L}_{i,\,j} \right)}{\left( 1 + x_{i,j}\, {\cal L}_{i,\,j} \right) \left( 1 + x_{i+1,j+1}\, {\cal L}_{i,\,j} \right)}
  \right]^{{\cal S}_{i,\,j}} \right\} \non
\eea
In  general, this expression is not zero as long as the distances
$x_{i,\,j}$ are arbitrary. However they are not all independent, being the
diagonals of a polygon in three spacetime dimensions. 
In Section \ref{vanishing} we come back to this result
and prove that it is actually zero when implementing an explicit
parametrization which constrains the $x_{i,\,j}$ segments
to be the diagonals of a three dimensional polygon.

\section{Connection with light--like Wilson loops}

In this Section we discuss the relation between the $n$--point
correlation function just computed and light--like polygonal Wilson
loops.

For the set of theories described by the action (\ref{action}) we
consider the Wilson loop operator
\begin{equation}
  \langle W_n(A, \hat{A}) \rangle = {\Big{\langle}} \, \frac{1}{2N}\,  Tr \, {\cal P} {\rm exp} \,\left( i\oint_{\G_n} A_\mu dz^\mu \right)
  +  \frac{1}{2M}\, Tr \, {\cal P} {\rm exp}  \,\left( i \oint_{\G_n} \hat{A}_\mu dz^\mu \right) {\Big{\rangle}}
\end{equation}
where $\G_n$ is a $n$--polygon with vertices $x_i, \, i=1, \cdots, n$,
and light--like edges, $x_{i,i+1}^2 = 0$.  We require all the
diagonals to be strictly positive in order to get real results. The
edges can be parametrized as
\begin{equation}
  z_i^\mu (\t_i) = x_i^\mu - x^\mu_{i,i+1} \, \t_i \qquad , \qquad 0 \leq \t_i \leq 1
\end{equation}
The perturbative evaluation of these operators up to two loops has
been carried on in \cite{HPW}. Here, we briefly summarize their
findings by pointing out what is needed for a comparison with
correlation functions.

The one--loop contribution to a WL is obtained by expanding the
path--ordered exponential at second order in the gauge
fields. Concentrating on one of the gauge fields, let's say $A_\mu$,
it is given by
\begin{equation}
  \label{WL1}
  \langle W(A)\rangle^{1-loop} = \frac{i^2}{N}\, \sum_{i \geq j} \int \, d\t_i\, d\t_j \, \dot{z}_i^\mu\, \dot{z}_j^\nu \,
  \langle Tr ( A_\mu(z_i)\, A_\nu(z_j)) \rangle
\end{equation}
where $\t_i, \t_j, i \neq j$ run independently between $0$ and $1$,
whereas for $i=j$ the integration domain is meant to be $0 \leq \t_i
\leq 1$ and $0 \leq \t_j \leq \t_i$.  Dots indicate derivatives with
respect to the affine parameters.

Plugging in the explicit expression for the gauge propagator, which in
Landau gauge reads
\begin{equation}
  \langle  (A_\mu)^a_{\, b}(z_i)\, (A_\nu)^c_{\, d}(z_j) \rangle =   -\frac{1}{8\pi K_1}\,
  \e_{\mu\nu\rho}\, \frac{(z_i - z_j)^\rho}{| z_i - z_j|^3} \, \d^a_d \, \d^c_b
\end{equation}
the contribution from a diagram where the gauge vector connects the
$(x_i,x_{i+1})$ and $(x_j,x_{j+1})$ edges is proportional to
$\e_{\mu\nu\rho}\, x_{i,i+1}^\mu\, x_{i+1,j}^\nu\, x_{j,j+1}^\rho\,
{\cal K}(i,j)$, where
\begin{eqnarray}
  && {\cal K}(i,j) = \frac{\pi^4}{2} \, \int_0^1 d\t_i d\t_j \, \times
  \\
  && \frac{1}{\left[(1-\t_i)\, (1-\t_j)\, x_{i,j}^2 + \t_i\, \t_j\, x_{i+1,j+1}^2 +
      \t_j\, (1-\t_i)\, x_{i,j+1}^2 + \t_i\, (1-\t_j)\, x_{i+1,j}^2 \right]^{\frac32}}
  \non
\end{eqnarray}
where we have taken into account that the contributions for $j=i$ and
$j=i+1$ vanish, due to the antisymmetry of the $\e$ tensor.

Now, including all the coefficients and summing the
analogous contribution coming from $\hat{A}$, the one--loop WL can be
written as
\begin{equation}
  \label{WL2}
  \langle W(A, \hat{A})\rangle^{1-loop} =  -\frac{1}{4\pi^5} \, \left( \frac{N}{K_1} + \frac{M}{K_2} \right) \,
  \sum_{i =1}^{n-2} \, \sum_{j= i+2}^{n-\d_{i,1}} \,  \e_{\mu\nu\rho}\, x_{i,i+1}^\mu\, x_{i+1,j}^\nu\, x_{j,j+1}^\rho \, {\cal K}(i,j)
\end{equation}
where the sum runs over all possible ways to connect two non--adjacent
lines.  We note that at this order matter fields do not enter the
calculation. Therefore, this result is valid also for pure
Chern--Simons theories.

As for the correlation functions, the overall color factor in
(\ref{WL2}) vanishes for all the theories with $K_2=-K_1$ and $M=N$,
ABJM case included. For this set of theories the correlation
functions/WL duality is then trivial at the first perturbative order.

Interesting non--trivial results can be found, instead, for theories
where the color factor does not vanish.  In fact, the main observation
is that, identifying the affine parameters $\t_i, \t_j$ with the
Feynman parameters $\b_1, \b_2$ in (\ref{5dint3}), the ${\cal K}(i,j)$
integral is precisely the same as the integral ${\cal J}(i,j)$ arising
in the computation of an $n$--point correlation function in the
light--like limit.  Since the integral (\ref{5dint3}) is the Feynman
parametrization of a 5d box integral, we can claim that also the
one--loop WL can be formally expressed in terms of a 5d scalar
integral.

The exact relation between correlation functions and WL, at one--loop
reads
\begin{equation}
  \label{identity}
  \lim_{x_{i,i+1}^2 \to 0} \frac{{\cal C}_n^{~1-loop}}{{\cal C}_n^{tree}} ~=~ \langle W(A, \hat{A})\rangle^{1-loop}
\end{equation}
all in terms of the 5d integral (\ref{5easy}).

We note that the two expressions coincide, independently of the values
of the couplings $K_1, K_2$ and for any value of the gauge ranks
$(N,M)$, as no planar limit is required.

\section{One--loop vanishing of correlators and Wilson loops}
\label{vanishing}

In this Section we give an analytical proof that  
the expression (\ref{finalresult2}) vanishes for any value of $n$. In
other words, the light--like limit of $n$--point correlation
functions of dimension--one BPS operators is zero at one loop.

Given the identification (\ref{identity}), as a by--product we also
prove that light--like $n$--polygon Wilson loops vanish at first
order. This result generalizes the one in \cite{HPW} valid only for
$n=4,\, 6$ and proves the conjecture made there that WL should be
one--loop vanishing for any $n$.

As we read in (\ref{finalresult2}), the one--loop correction to a correlation function
is proportional to the logarithm of a product of factors with
schematic form $\left(\frac{1\pm x\, {\cal L}_{i,\,j} }{1\mp x\, {\cal
    L}_{i,\,j}}\right)^{{\cal S}_{i,\,j}}$.  We prove  that
this product always evaluates to 1.

 In (\ref{finalresult2}) the factors are grouped according to the pair of edges involved in a
given gauge vector exchange (see blocks in Fig. \ref{fig:buildingblock}).  The basic idea of the proof is to
reorganize them by group together all the factors which 
depend on the same diagonal $x_{i,j}$. It is easy to ascertain
that each long diagonal is involved in four contributions, coming from
the four possible interactions connecting the edges which are adjacent
to the diagonal itself (See Fig.\ref{fig:genericBlock} (a)). In the
case of a short diagonal, one of these contributions vanishes 
(it would be a correction to the vertex), thus we are left with just
three pieces (See Fig.\ref{fig:genericBlock}(b)).

\FIGURE{
  \centering
  \includegraphics[width=0.9\textwidth]{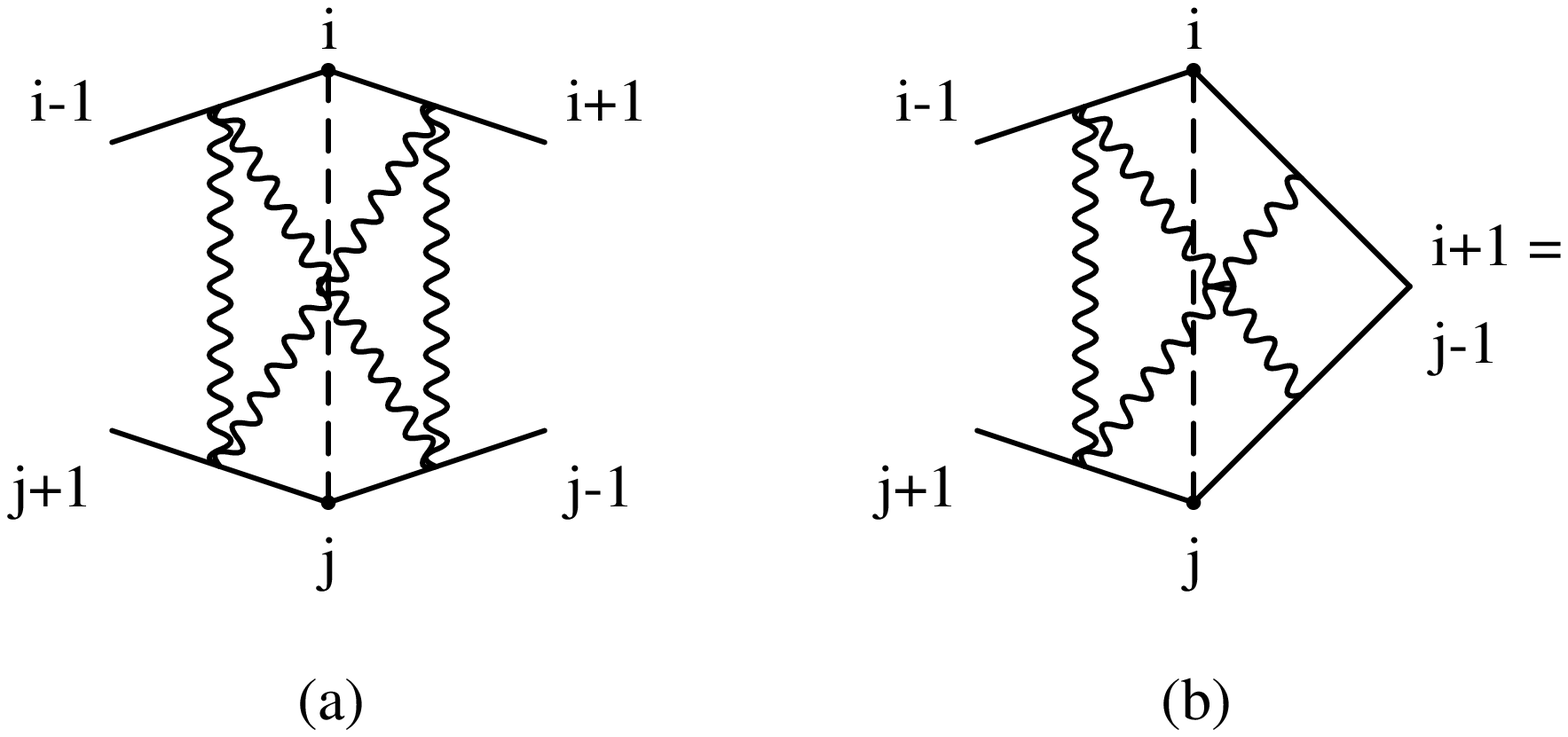}
  \caption{In picture (a) the four blocks in which the
    reference diagonal $x_{i,j}$ is involved are depicted. In picture
    (b) the case of a short diagonal and its three blocks is
    shown. Each wiggled line has to be interpreted separately.}
  \label{fig:genericBlock}
}

Once this reshuffling of factors has been performed in (\ref{finalresult2}), we
prove that the product of contributions involving the same reference diagonal
evaluates to $+1$ for long diagonals and to $-1$ for short ones. 
We consider 
a generic diagonal and parametrize all distances in full generality, so that once we 
establish this property for one diagonal, we can apply it to all the
contributions to the correlator.

Let us focus on one particular diagonal $x_{i,j}$, and suppose it is
long.  The corresponding block of factors then depends only on the
nearest neighbours of the vertices $x_i$ and $x_j$, which are
$x_{i-1}$, $x_{i+1}$, $x_{j-1}$, $x_{j+1}$. These six points are
parametrized by 18 coordinates. However, four of them can be eliminated by
light--likeness of the edges $x_{i,\,i+1}$, $x_{i,\,i-1}$,
$x_{j,\,j+1}$, $x_{j,\,j-1}$. By using translation invariance, we choose a convenient 
reference frame where $x_{i}^\mu=(0,0,0)$, so removing three more coordinates.
Using rotational invariance, we eliminate two further parameters by choosing 
$x_{j}^\mu = (0,b,0)$ where $b>0$. In this way, the reference diagonal lies in the $t=0$ plane.
We parametrize the rest of the block in terms of the nine remaining variables as
follows
\begin{eqnarray}
  \label{param}
  x_{i-1}^\mu &=& r_1\left(1,\cos{\f_1} ,\sin{\f_1 } \right),\
  \ \ \ \ \ \ \ \ x_{i+1}^\mu = r_3\left(1,\cos{\f_3} ,\sin{\f_3} \right)\non\\
  x_{j-1}^\mu &=& x_{j}^\mu+r_2\left(1,\cos{\f_2} ,\sin{\f_2} \right),\
  \ x_{j+1}^\mu = x_{j}^\mu+r_4\left(1,\cos{\f_4} ,\sin{\f_4} \right)
\end{eqnarray}
This parametrization is sketched in Fig. \ref{fig:parametrization}: the
$\phi_i$'s are the angles held by the projections of the light--like
lines on the $t=0$ plane, while the moduli of the $r_i$'s measure the
lengths of these same projections.
\FIGURE{ \centering
  \includegraphics[width=0.8\textwidth]{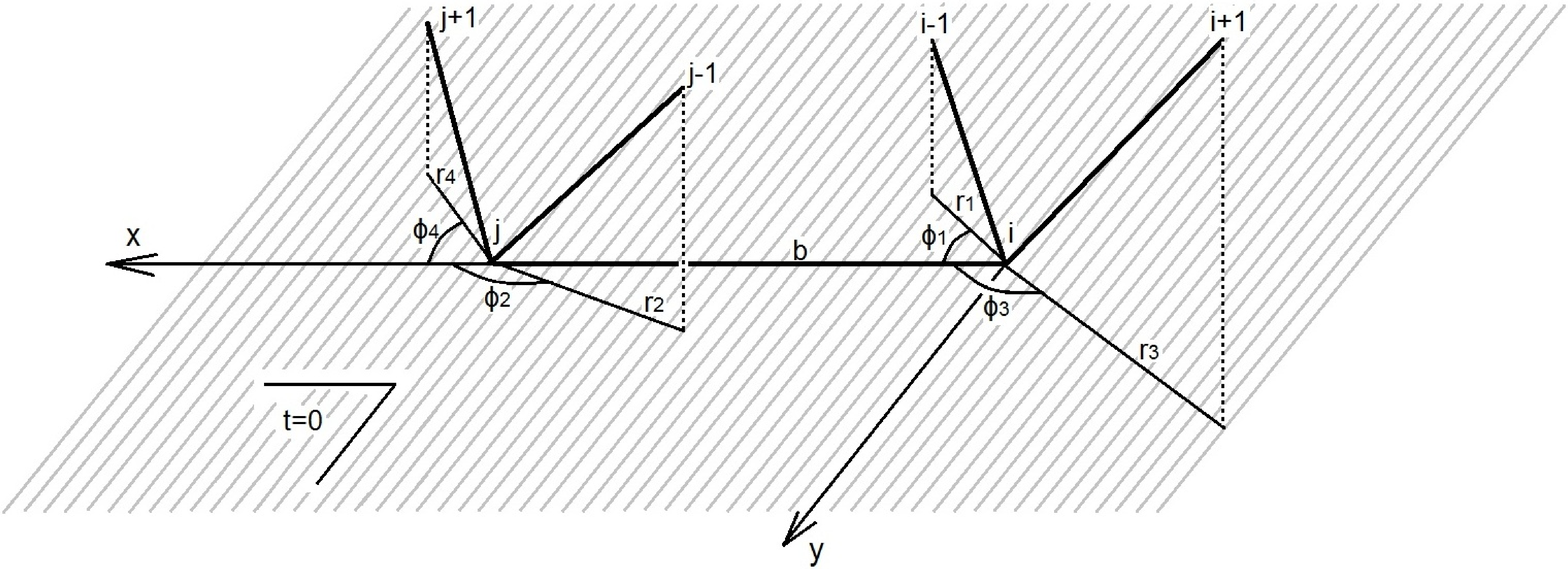}
  \caption{Parametrization of the block of contributions involving the
    same reference diagonal $x_{i,\,j}$.}
  \label{fig:parametrization}
}
It is obvious that the edges are light--like and the reference diagonal
$x_{i,\,j}$ is space--like by construction. At this stage, the other diagonals are 
not necessarily space--like. The quest for them to be space--like implies that
$r_1$, $r_3$ and $r_2$, $r_4$ should have
separately the same sign, in order for adjacent segments to point
alternatively to the future and to the past. In the following we will
assume that they are all positive, but the final statement can be
exhaustively shown to be valid for any choice of these signs.

Let us now evaluate the product of the four contributions for the
reference diagonal $x_{i,\,j}$, namely
\begin{eqnarray}
\label{long}
  && \left( \frac{1+x_{i,j}\,\,{\cal L}_{i,\,j}}{1-x_{i,j}\,\,{\cal L}_{i,\,j}} \right)^{\mathcal{S}_{i,\,j}} ~
  \left( \frac{1+x_{i,j}\,\,{\cal L}_{i-1,\,j-1}}{1-x_{i,j}\,\,{\cal L}_{i-1,\,j-1}} \right)^{\mathcal{S}_{i-1,\, j-1}} \non\\ &&
  \left( \frac{1-x_{i,j}\,\,{\cal L}_{i-1,\,j}}{1+x_{i,j}\,\,{\cal L}_{i-1,\,j}} \right)^{\mathcal{S}_{i-1,\, j}} ~
  \left( \frac{1-x_{i,j}\,\,{\cal L}_{i,\,j-1}}{1+x_{i,j}\,\,{\cal L}_{i,\,j-1}} \right)^{\mathcal{S}_{i,\,j-1}}
\end{eqnarray}
By plugging in the parametrization (\ref{param}) we obtain a nice symmetric expression
\begin{eqnarray}\label{parametrized}
  && \left(\frac{1+\left|\frac{\sin{\left(\frac{\f_1 -\f_2  }{2}\right)}}{\cos{\left(\frac{\f_1 +\f_2 }{2}\right)}}
      \right|}{1-\left|\frac{\sin{\left(\frac{\f_1 -\f_2  }{2}\right)}}{\cos{\left(\frac{\f_1 +\f_2 }{2}\right)}}
      \right|}\right)^{\text{Sign}\left[\frac{\sin{\left(\frac{\f_1 -\f_2  }{2}\right)}}{\cos{\left(\frac{\f_1 +\f_2 }{2}\right)}}
    \right]}
  \left(\frac{1-\left|\frac{\sin{\left(\frac{\f_3 -\f_2 }{2}\right)}}{\cos{\left(\frac{\f_3 +\f_2 }{2}\right)}}
      \right|}{1+\left|\frac{\sin{\left(\frac{\f_3 -\f_2 }{2}\right)}}{\cos{\left(\frac{\f_3 +\f_2 }{2}\right)}}
      \right|}\right)^{\text{Sign}\left[\frac{\sin{\left(\frac{\f_3 -\f_2 }{2}\right)}}{\cos{\left(\frac{\f_3 +\f_2 }{2}\right)}}
    \right]} \non\\ &&
  \left(\frac{1-\left|\frac{\sin{\left(\frac{\f_1 -\f_4 }{2}\right)}}{\cos{\left(\frac{\f_1 +\f_4 }{2}\right)}}
      \right|}{1+\left|\frac{\sin{\left(\frac{\f_1 -\f_4 }{2}\right)}}{\cos{\left(\frac{\f_1 +\f_4 }{2}\right)}}
      \right|}\right)^{\text{Sign}\left[\frac{\sin{\left(\frac{\f_1 -\f_4 }{2}\right)}}{\cos{\left(\frac{\f_1 +\f_4 }{2}\right)}}
    \right]}
  \left(\frac{1+\left|\frac{\sin{\left(\frac{\f_3 -\f_4 }{2}\right)}}{\cos{\left(\frac{\f_3 +\f_4 }{2}\right)}}
      \right|}{1-\left|\frac{\sin{\left(\frac{\f_3 -\f_4 }{2}\right)}}{\cos{\left(\frac{\f_3 +\f_4 }{2}\right)}}
      \right|}\right)^{\text{Sign}\left[\frac{\sin{\left(\frac{\f_3 -\f_4 }{2}\right)}}{\cos{\left(\frac{\f_3 +\f_4 }{2}\right)}}
    \right]}
\end{eqnarray}
where $\text{Sign}(x)$ is the sign function. We notice that the
explicit parametrization allows us to fix a loose end from Section
\ref{sec:oneloop}, namely we have ascertained that the terms
$x_{i,j}\,{\cal L}_{\dots}$ are real positive functions. Furthermore we
observe that the apparently awkward exponents ${\cal S}_{i,\,j}$
(\ref{signs_definition}) are surprisingly just $\pm$
signs.\\ Expression (\ref{parametrized}) can be written in a compact
fashion (here and in the following $\f_5 = \f_1$ is understood)
\begin{eqnarray}\label{product}
  \prod_{i=1}^4\,\,  \left(\frac{1+\left|\frac{\sin{\left(\frac{\f_i -\f_{i+1}  }{2}\right)}}{\cos{\left(\frac{\f_i +\f_{i+1} }{2}\right)}}
      \right|}{1-\left|\frac{\sin{\left(\frac{\f_i -\f_{i+1}  }{2}\right)}}{\cos{\left(\frac{\f_i +\f_{i+1} }{2}\right)}}
      \right|}\right)^{\text{Sign}\left[\frac{\sin{\left(\frac{\f_i -\f_{i+1}  }{2}\right)}}{\cos{\left(\frac{\f_i +\f_{i+1} }{2}\right)}}
    \right]}
\end{eqnarray}
We observe that the expression depends exclusively on the four angles of
the parametrization but not on any of the five dimensionful
parameters. We also note that in each contribution the arguments of
absolute values and $\text{Sign}$ functions are the same. Because of
that and using the fact that $\left( \frac{1 \pm x}{1 \mp x}
\right)^{\pm} = \frac{1 + x}{1 - x}$ we may simplify expression (\ref{product}) to
obtain
\begin{eqnarray}
  \prod_{i=1}^4\,\, \frac{\cos{\left(\frac{\f_i +\f_{i+1} }{2}\right)} + \sin{\left(\frac{\f_i -\f_{i+1}  }{2}\right)}}{\cos{\left(\frac{\f_i +\f_{i+1} }{2}\right)} - \sin{\left(\frac{\f_i -\f_{i+1}  }{2}\right)}}
\end{eqnarray}

This is equivalent to
\begin{eqnarray}
  \prod_{i=1}^4\,\, \cot{\left(\frac{\f_i}{2} - \frac{\p}{4}\right)}\,\, \tan{\left(\frac{\f_{i+1}}{2} - \frac{\p}{4}\right)} = 1
\end{eqnarray}
Therefore, this completes the proof for long diagonals.

For a short diagonal $x_{i,\,j}$, it suffices to take the result above 
and set e.g., $x_{i+1} = x_{j-1}$. Then the contribution involving
${\cal L}_{i,\,j-1}$ vanishes by construction leaving
 \begin{equation}\label{short}
  \frac{\cos{\left(\frac{\f_1 +\f_2 }{2}\right)}+\sin{\left(\frac{\f_1 -\f_2  }{2}\right)}}{\cos{\left(\frac{\f_1 +\f_2 }{2}\right)}-\sin{\left(\frac{\f_1 -\f_2  }{2}\right)}} ~
  \frac{\cos{\left(\frac{\f_1 +\f_4 }{2}\right)}-\sin{\left(\frac{\f_1 -\f_4  }{2}\right)}}{\cos{\left(\frac{\f_1 +\f_4 }{2}\right)}+\sin{\left(\frac{\f_1 -\f_4  }{2}\right)}} ~
  \frac{\cos{\left(\frac{\f_3 +\f_4 }{2}\right)}+\sin{\left(\frac{\f_3 -\f_4  }{2}\right)}}{\cos{\left(\frac{\f_3 +\f_4 }{2}\right)}-\sin{\left(\frac{\f_3 -\f_4  }{2}\right)}} ~
\end{equation}
When parametrizing as in eq. (\ref{param}) the condition $x_{i+1} = x_{j-1}$ is forced by choosing 
\begin{eqnarray}
  r_2 = r_3, \quad
  r_3\, \cos{\left(\f_3 \right)}= b+r_2\, \cos{\left(\f_2 \right)}, \quad
  \sin{\left(\f_3 \right)} = \sin{\left(\f_2 \right)}
\end{eqnarray}
These equations are solved by $\phi_3 = \pi - \phi_2$ and some function $r_2=r_2(a,\phi_3)$ which is irrelevant. 
Plugging it into (\ref{short}) finally simplifies the expression to $-1$, in a
completely analogous way as in the long diagonal case. Since there are $n$
contributions of the short type, and since $n$ is even, the overall
contribution of short diagonals is equal to $+1$.

Summarizing, we have shown that the combined collection of all short
and long diagonal contributions to the argument of the logarithm is
equal to $+1$. Therefore, the logarithm is equal to zero, thus proving
the vanishing of the $n$--point correlator and Wilson loops at one loop.

\section{Generalization to higher dimensional operators}
\label{higher}

So far we have considered one--loop corrections $\mathcal{C}_n^{1-loop}$ to
correlators of dimension--one operators ${\cal O}^{i}_{j} = \Tr (A^{i}
B_{j})$. In this Section, we show that one--loop corrections to
correlators $\mathcal{C}_{n,2l}$ of higher dimensional operators
(\ref{BPS}) can be simply computed once $\mathcal{C}_n^{1-loop}$ is
known. In particular, since $\mathcal{C}_n^{1-loop}$ is zero, the same
holds for any correlator $\mathcal{C}_{n,2l}$ with $l>1$. We emphasize
that the derivation of this result is valid for any value of the gauge
group parameters $(N,M)$.

The most divergent part of connected correlators of
higher dimensional operators in the light--like limit
$x^2_{i,i+1}\rightarrow 0$ at tree level reads
\begin{eqnarray}
  \label{tree}
  &&\mathcal{C}^{tree}_{n,2l} ~\propto ~  \sum^{2l-1}_{s=1}  \mathcal{T}^{tree}_s
  \nonumber \\
  &&\mathcal{T}^{tree}_s = \prod_{j=1}^{n/2}
  \left(\frac{1}{x_{2j-1,2j}} \right)^s \left(\frac{1}{x_{2j,2j+1}} \right)^{2l-s}
\end{eqnarray}
Eq. (\ref{tree}) extends eq. (\ref{cyclic}) to the $l>1$ case.  The
general contribution in the sum (\ref{tree}) is a polygon with edges
alternately made by $s$ and $2l-s$ propagators (see
Fig. \ref{figura1}(a)). Each value of $s$ defines a different
topology $\mathcal{T}_s$. In the rest of the discussion, it is useful to divide 
each topology $\mathcal{T}_s$ into classes
$\mathcal{T}_{s,a}$ where the parameter $a$ counts the number of
$\langle A \Ab \rangle$ propagators inside a block of $s$ lines (see Fig. \ref{figura1}(b)).

\FIGURE{
  \centering
  \label{figura1}
  $
  \begin{array}{ccc}
   \begin{array}{c}
    \raisebox{-9.5 ex}{\includegraphics[width=0.33\textwidth,angle=0]{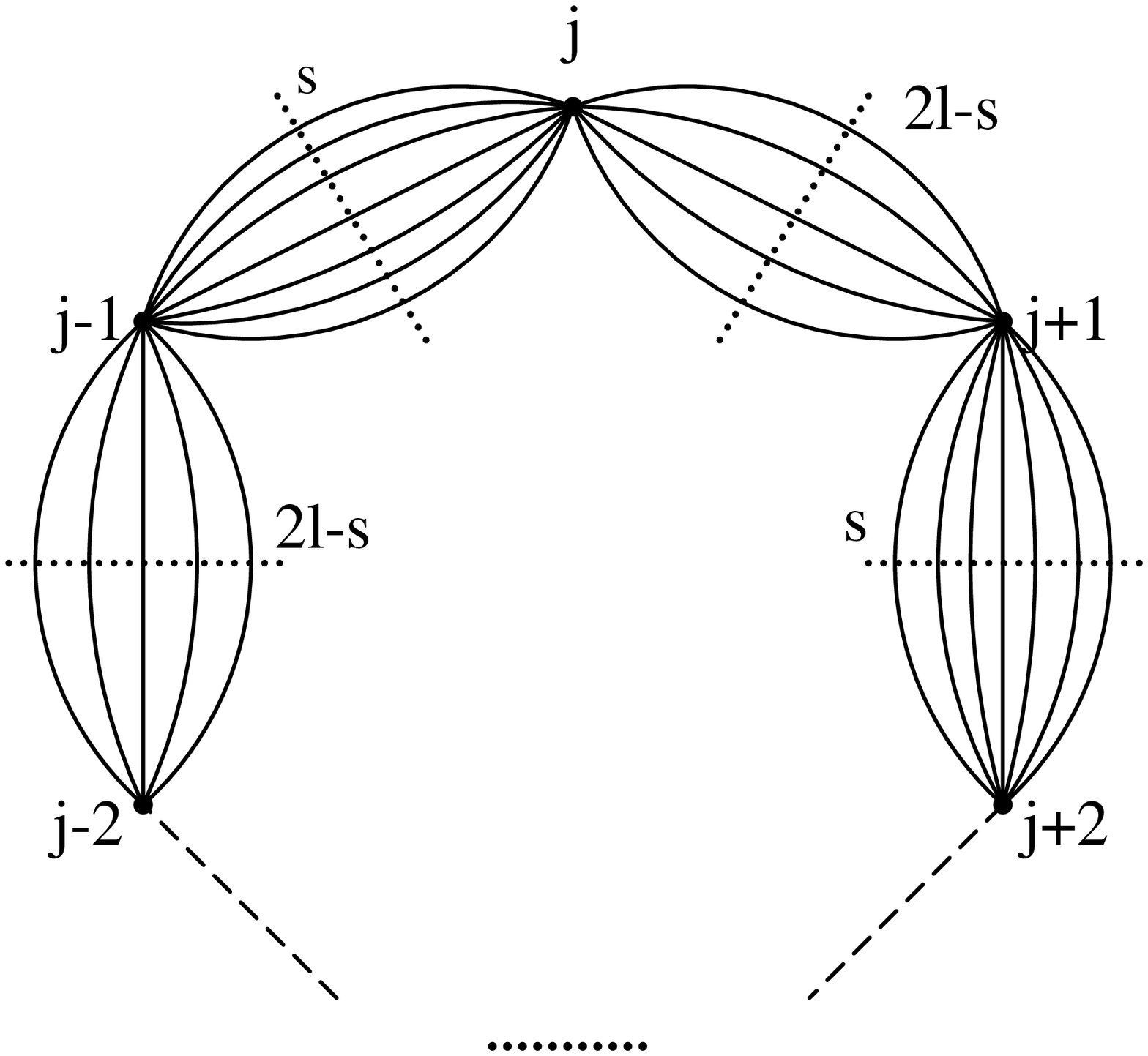}} \\ \\
    \bf{(a)}
  \end{array}
    &
    \phantom{a}
    \hspace{10mm}
    &
   \begin{array}{c}
    \phantom{a} \\
    \raisebox{-6.5 ex}{\includegraphics[width=0.45\textwidth,angle=0]{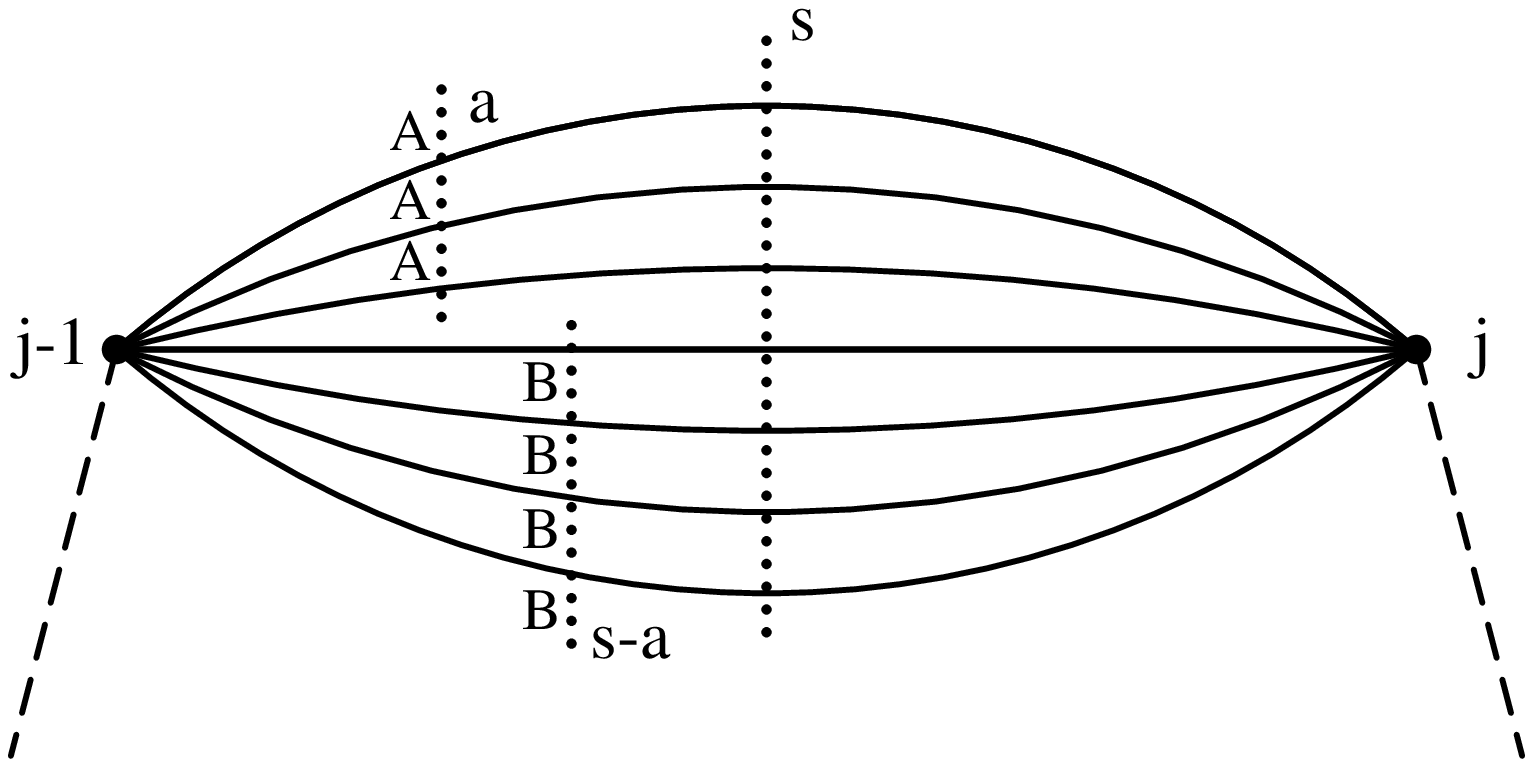}} \\ \\ \\
    \bf{(b)}
  \end{array}  
  \end{array}
  $
  \caption{General form of the contributions to
    $\mathcal{C}_{n,2l}^{tree}$. In Fig.  (a), structure of the
    leading divergent terms in the limit $x_{i,i+1}^2\rightarrow
    0$. In Fig.  (b), the parameter $a$ counts the number of
    $\langle A \Ab \rangle$ propagators in a set of $s$ lines.}
}

One--loop corrections to $\mathcal{C}_{n,2l}$ are given by inserting a
gauge propagator $V$ or $\hat{V}$ in all possible ways between the
edges of the polygon $\mathcal{C}_{n,2l}^{tree}$.

As in the $l=1$ case, the only non--trivial insertions occur when the
gauge propagator connects  two
non--consecutive edges in the polygon. All other possible insertions
are zero due to D--algebra constraints or to the antisymmetry of the
$\epsilon$ tensor.

The non--trivial corrections have the form (\ref{block}). However, since
now we have more than one chiral propagator in each edge, we have more
than one possibility to insert a gauge line between the same two edges
of a correlator.

The combinatorial factor is in principle different for corrections
involving different pairs of edges in each class
$\mathcal{T}_{s,a}$. However, a careful computation which takes into
account the relative signs between $A$ and $B$ vertices
(\ref{vertices}) and between the two building blocks (\ref{block})
shows that the combinatorial factor depends only on the $(a,s)$ parameters 
and it is thus a common factor for all corrections inside each
$\mathcal{T}_{s,a}$ class. Precisely, the one--loop correction to the  
generic $\mathcal{T}_{s,a}$ class reads
\begin{eqnarray}
  \label{subtopologies}
  \mathcal{T}_{s,a}^{1-loop} \propto \mathcal{T}_{s,a}^{tree} \times (s-2a)^2 ~\, \sum_{i=1}^{n-2}\,\, \sum_{j=i+2}^{n-\d_{i,1}} \,\e_{\m\n\rho}\, x_{i,\,i+1}^{\m}\, x_{i+1,\,j}^{\n}\, x_{j,\,j+1}^{\rho}\, {\cal J}(i,j)
\end{eqnarray}
where ${\cal J}(i,j)$ is the five dimensional box integral
(\ref{5easy}).

This formula closely resembles eq. (\ref{sum}). In particular, the sums
in these two expressions are the same. Thus, having computed the
one--loop corrections to the $n$--point function for dimension--one
operators, we  immediately have the result for any $\mathcal{T}_{s,a}^{1-loop}$. 
The complete one--loop correction to the correlator
$\mathcal{C}_{n,2l}$ can be then recovered through (\ref{tree}). 
In particular, since we have proved that $\mathcal{C}_{n}^{1-loop}/\mathcal{C}_{n}^{tree}$ vanishes in the light--like limit, so 
$\mathcal{C}_{n,2l}^{1-loop}/\mathcal{C}_{n,2l}^{tree}$ does.

\section{Conclusions}

In this paper we have focused on the novel proposal that a multiple
light--like limit of the correlation function of $n$ protected
operators reproduces a Wilson loop evaluated on a null
$n$-polygon. This statement has been argued and verified
perturbatively in ${\cal N}=4$ SYM, and it has been claimed to be
valid for any conformal gauge theory in any dimensions
\cite{AEKMS}.

We have confirmed this expectation for a class of
supersymmetric Chern--Simons matter theories in three dimensions at
first order in perturbation theory.  Our check goes as follows: We
have computed one--loop corrections to the correlation function of $n$
BPS scalar operators in a manifestly ${\cal N}=2$ supersymmetric
formalism. Remarkably, they can be expressed in terms of five
dimensional box integrals.  Then we have performed the light--like
limit of the correlator as prescribed in \cite{AEKMS} in order to
compare it to the Wilson loop expectation values on $n$-cusped
light--like polygons. These were found explicitly in \cite{HPW} in the
cases of $n=4$ and $n=6$ cusps. The former was shown to vanish
analytically, whereas numerical evidence hinted at the vanishing of the
latter. This suggested that all light--like Wilson loop should not
receive first order quantum corrections in Chern--Simons theories.  We
have managed to show analytically that both one--loop corrections to
the correlators and to the Wilson loops vanish, thus confirming the
correlator/Wilson loop duality for a class of three dimensional
theories, and proving the claim on the vanishing of light--like Wilson
loops at first order.  We point out that our check is not just a mere
identity between two vanishing contributions, since the equality
between correlators and Wilson loops in the light--like limit already
holds when expressing them in terms of integrals, before showing that
these expressions actually vanish.

In our computation this
equivalence seems to apply to any ${\cal N}=2$ Chern--Simons matter
theories, but this is just an artifact of the low perturbative
order. Indeed quantum corrections arise purely from the Chern--Simons
sector both for the correlation functions and for Wilson loops and the
matter sector is not involved. On one hand, this confirms the idea
that the relation should be valid in any conformal field theory:
Indeed, all Chern--Simons matter theories are naively conformal
invariant at one loop.  On the other hand, this shadows any difference
between the gauge theories spanned by our ${\cal N}=2$ Lagrangian
(\ref{action}), both as concerns supersymmetry and
conformality. Models with different amounts of symmetry should be
discriminated starting from two--loop order, where we expect that the
equivalence between correlators and Wilson loops may hold for the
subset of conformal field theories only.  The Wilson loop on a four
cusped null contour is available at two loops in literature
\cite{HPW}; the computation of correlation functions at the same order
is then highly desirable and is planned for a future investigation
\cite{preparation1}.\\

The correlators/Wilson loop equivalence is just a corner of the chain
of dualities conjectured in \cite{AEKMS,EKS}. Dualities involving
scattering amplitudes are of great interest, the hope being to
eventually extract information on those from the knowledge of simpler
objects such as Wilson loops and correlation functions.

In three dimensions results on loop amplitudes are limited to the scattering of
four external particles at first order \cite{ABM}. When the theory
possesses enough supersymmetry these amplitudes have been shown to
vanish, completing the test of dualities in the one--loop $n=4$
case. Differently from the correlator/Wilson loop equivalence,
dualities involving scattering amplitudes seem to require
supersymmetry already at one loop, indicating that their origin should
be different from the former. Indeed the duality between MHV
scattering amplitudes and Wilson loops is intimately connected to dual
superconformal invariance \cite{Drummond:2008vq} on the field theory
side and to T-duality in the AdS dual description
\cite{BM}. Results on dual superconformal invariance have been
extended to tree level scattering amplitudes in three dimensional
theories in \cite{HL, GHKLL}, whereas fermionic T-dualities seem to be
ill-defined for the $\s$--model in the dual picture
\cite{ADO}--\cite{DO}.\\ In order to shed more light on the role of
superconformal invariance and dualities for Chern--Simons matter
theories the knowledge of a larger sample of scattering amplitudes is
mandatory. In particular it would be highly desirable to compute the
six point amplitude at one--loop order and the four point amplitude at
two loops, which should not be trivial. This task represents another
challenging line of research \cite{preparation2}.

\vskip 25pt
\section*{Acknowledgements}
\noindent

This work has been supported in part by INFN and MIUR.

\vfill
\newpage
\appendix
\section{Notations and conventions}

For three dimensional ${\cal N}=2$ superspace we follow the conventions of \cite{superspace}.\\
The metric for the fermionic coordinates $\th^{\a}$ ($\a = 1,\, 2$) of
${\cal N} = 2$ superspace is
\begin{eqnarray}
  C^{\alpha\beta} = \left(\begin{array}{cc} 0 & i \\ -i & 0 \end{array}\right) \qquad
  C_{\alpha\beta} = \left(\begin{array}{cc} 0 & -i \\ i & 0 \end{array}\right)
\end{eqnarray}
which is used to rise and lower spinorial indices as
\begin{eqnarray}
  \psi^\alpha=C^{\alpha\beta}\psi_\beta  \qquad \psi_\alpha=\psi^\beta C_{\beta\alpha}
\end{eqnarray}
and obeys the relation
\begin{eqnarray}
  C^{\alpha\beta}\, C_{\gamma\delta}
  &=\delta^\alpha{}_\gamma\, \delta^\beta{}_\delta - \delta^\alpha{}_\delta\, \delta^\beta{}_\gamma
\end{eqnarray}
Spinors are contracted according to
\begin{eqnarray}
  \psi\chi=\psi^\alpha\, \chi_\alpha=\chi^\alpha\, \psi_\beta=\chi\psi
  \qquad
  \psi^2=\frac{1}{2}\, \psi^\alpha\, \psi_\alpha
\end{eqnarray}
We consider a three dimensional Minkowski spacetime with mostly plus
signature $g^{\mu\nu}=\mathrm{diag}\left(-1,1,1\right)$.  The
corresponding Dirac $\left(\g^{\mu}\right)^\alpha{}_\beta$ matrices
satisfy the algebra
\begin{eqnarray}
  (\gamma^\mu)^\alpha{}_\gamma\, (\gamma^\nu)^\gamma{}_\beta
  =-g^{\mu\nu}\delta^\alpha{}_\beta + i\, \epsilon^{\mu\nu\rho}\, (\gamma_\rho)^\alpha{}_\beta
\end{eqnarray}
The following identities for traces of Dirac matrices can be read from
the above algebra
\begin{eqnarray}
  &\tr(\gamma^\mu\, \gamma^\nu)
  &= (\gamma^\mu)^\alpha{}_\beta\, (\gamma^\nu)^\beta{}_\alpha
  = - 2\,  g^{\mu\nu} \\
  &\tr(\gamma^\mu\, \gamma^\nu\, \gamma^\rho)
  &= - (\gamma^\mu)^\alpha{}_\beta\, (\gamma^\nu)^\beta{}_\gamma\, (\gamma^\rho)^\gamma{}_\alpha
  = 2\,i\,\epsilon^{\mu\nu\rho} \\
  &\tr(\gamma^\mu\, \gamma^\nu\, \gamma^\rho\, \gamma^\sigma) &= (\gamma^\mu)^\alpha{}_\beta\, (\gamma^\nu)^\beta{}_\gamma\,
  (\gamma^\rho)^\gamma{}_\delta\, (\gamma^\sigma)^\delta{}_\alpha = \non\\
  &&= 2\, (g^{\mu\nu}\, g^{\rho\sigma}-g^{\mu\rho}\, g^{\nu\sigma}+g^{\mu\sigma}\, g^{\nu\rho})
\end{eqnarray}
The scalar product of two bispinors follows
\begin{eqnarray}
  p^{\alpha\beta}\,k_{\alpha\beta}\, =\, 2\, p\cdot k
\end{eqnarray}
Vectors and bispinors are exchanged according to
\begin{eqnarray}\begin{array}{lll}
    \mathrm{for~ coordinates}\qquad   & x^{\alpha\beta} = \frac{1}{2}\, (\gamma_\mu)^{\alpha\beta}\, x^\mu  &\qquad
    x^\mu=(\gamma^\mu)_{\alpha\beta}\, x^{\alpha\beta} \\
    \mathrm{for~ derivatives}\qquad    & \partial_{\alpha\beta}=(\gamma^\mu)_{\alpha\beta}\, \partial_\mu  & \qquad
    \partial_\mu=\frac{1}{2}\, (\gamma_\mu)^{\alpha\beta}\, \partial_{\alpha\beta} \\
    \mathrm{for~ fields}\qquad     &
    A_{\alpha\beta}=\frac{1}{\sqrt{2}}\, (\gamma^\mu)_{\alpha\beta}\, A_\mu  & \qquad
    A_\mu=\frac{1}{\sqrt{2}}\, (\gamma_\mu)^{\alpha\beta}\, A_{\alpha\beta}
\end{array}
\end{eqnarray}
Supercovariant derivatives are defined as
\begin{equation}
  D_\a = \pa_\a + \frac{i}{2}\,  \thb^\b\,  \pa_{\a\b}
  \qquad , \qquad \Db_\a = \bar\pa_\a
  + \frac{i}{2}\,  \th^\b\,  \pa_{\a\b}
\end{equation}
and satisfy the anticommutator
\begin{equation}
  \{D_\a ,\,  \Db_\b\} = i\,  \pa_{\a\b}
\end{equation}

The components of a chiral and an anti-chiral superfield,
$Z(x_L,\theta)$ and $\bar{Z}(x_R,\bar{\theta})$, are a complex boson
$\phi$, a complex two-component fermion $\psi$ and a complex auxiliary
scalar $F$. Their component expansions are given by
\begin{eqnarray}
  Z = \phi(x_L) + \th^\a \psi_\a(x_L) - \th^2 \,
  F(x_L) \non \\ \bar{Z} = \bar{\phi}(x_R)
  + \thb^\a \bar{\psi}_\a(x_R) - \thb^2 \, \bar{F}(x_R)
\end{eqnarray}
where $x_L^\mu = x^\mu + i \theta \gamma^\mu \bar{\theta}$, $x_R^\mu =
x^\mu - i \theta \gamma^\mu \bar{\theta}$.

The components of the real vector superfield
$V(x,\theta,\bar{\theta})$ in Wess-Zumino gauge ($V| = D_\a V| = D^2
V| = 0$) are the gauge field $A_{\a\b}$, a complex two-component
fermion $\lambda_\a$, a real scalar $\sigma$ and an auxiliary scalar
$D$, such that
\begin{equation}
  \label{eqn:WZgauge}
  V = i \, \th^\a \thb_\a \, \sigma(x)
  + \th^\a \thb^\b \, \sqrt{2} \, A_{\a\b}(x)
  - \th^2 \, \thb^\a \bar{\lambda}_\a(x)
  - \thb^2 \, \th^\a \lambda_\a(x)
  + \th^2 \, \thb^2 \, D(x)
\end{equation}

\bigskip
\noindent
The $U(N)$ generators are $T^A = (T^0, T^a)$, where $T^0 =
\frac{1}{\sqrt{N}}$ and $T^a$ ($a=1,\ldots, N^2-1$) are a set of
$N\times N$ hermitian matrices.  The generators are normalized as
$\Tr( T^A T^B )= \delta^{AB}$.

\section{The emergence of a five dimensional integral}
\label{AppB}

In this appendix we give a detailed proof of eq. (\ref{5dint}) which
allows to express a double three dimensional integral as a one--loop
five dimensional box integral.

In order to simplify the notation, in the expression
$\e_{\m\n\rho}\, \partial_{i}^{\m}\, \partial_{i+1}^{\n}\, \partial_{j+1}^{\rho}\,
{\cal I}(i,j) $ we choose $i=1, j=3$.  Applying the derivatives to the
integrand, the expression that we need evaluate is then
\begin{eqnarray}
  \label{initial}
  &&\epsilon^{\mu\nu\rho}\,\partial_{1\mu}\,\partial_{2\nu}\,\partial_{4\rho}\,
  \int d^{3}x_0\,d^3x_5\,\frac{1}{x_{1,0}\,x_{2,0}\,x_{0,5}\,x_{3,5}\,x_{4,5}}
  \\
  \non \\
  &~&=  -\epsilon_{\mu\nu\rho} \, \int d^{3}x_0\,d^3x_5\,\frac{x_{1,0}^\mu\,x_{2,0}^{\nu}\,x_{4,5}^{\rho}}
  {(x_{1,0}^2)^{3/2}\,(x_{2,0}^2)^{3/2}\,(x_{0,5}^2)^{1/2}\,(x_{3,5}^2)^{1/2}\,(x_{4,5}^2)^{3/2}} \equiv  \, \mathcal{I}
  \non
\end{eqnarray}
We first focus on the $x_0$--integration which can be performed by
introducing Feynman parameters
\begin{eqnarray}
  \label{int_x5}
  &&\epsilon_{\mu\nu\rho}\int d^3x_0 \,
  \frac{x_{1,0}^\mu\,x_{2,0}^{\nu}}
  {(x_{1,0}^2)^{3/2}\,(x_{2,0}^2)^{3/2}\,(x_{0,5}^2)^{1/2}}=\nonumber\\
  &&\frac{4}{\pi^{3/2}}\, \Gamma\left( \frac{7}{2}\right) \int \prod\limits_{i=1}^3dy_i\,\delta(\sum y_i-1)y_1^{1/2}y_2^{1/2}y_3^{-1/2}
  \int d^3x_0 \frac{\epsilon_{\mu\nu\rho}\,x_{1,0}^\mu\,x_{2,0}^{\nu}}{[(x_0-\rho_1)^2+\Omega_1]^{7/2}}
\end{eqnarray}
where $\rho_1^\mu=y_1 x_1^\mu+y_2 x_2^\mu+y_3 x_5^\mu$ and
$\Omega_1=y_1 y_2 x_{1,2}^2+y_1 y_3 x_{1,5}^2+y_2 y_3 x_{2,5}^2$.

Performing the shift $x_0^\mu\to x_0^\mu+\rho_1^\mu$ and integrating
over $x_0$ we obtain
\begin{equation}
  4\,\epsilon_{\mu\nu\rho}\,x_{1,5}^\mu\, x_{2,5}^\nu
  \int \prod\limits_{i=1}^3dy_i\,\delta(\sum_i y_i-1)\frac{(y_1\,y_2\,y_3)^{1/2}}
  {(y_1 y_2 x_{1,2}^2+y_1 y_3 x_{1,5}^2+y_2 y_3 x_{2,5}^2)^2}
  \label{x0_int}
\end{equation}
Now, in order to render the remaining $x_5$ integration in
(\ref{initial}) doable, we manipulate the expression (\ref{x0_int}) by
using the Mellin-Barnes integral representation. According to the
general identity
\begin{equation}
  \frac{1}{(k^2 + A^2 + B^2)^a} = \frac{1}{(k^2)^a \G(a)} \, \int_{-i\infty}^{+i\infty} \frac{ds dt}{(2\pi i )^2} \G(-s) \G(-t) \G(a +s+t)
  \left( \frac{A^2}{k^2} \right)^s \left( \frac{B^2}{k^2} \right)^t
  \label{MB}
\end{equation}
we rewrite (\ref{x0_int}) as
\begin{equation}
  \frac{4\,\epsilon_{\mu\nu\rho}\,x_{1,5}^\mu\, x_{2,5}^\nu}{\sqrt{\pi}}
  \int\limits_{-i\infty}^{i\infty}\frac{du\,dv}{(2\pi i)^2}
  \frac{\Gamma\left(-u|-\frac{1}{2}-u|-v|-\frac{1}{2}-v|2+u+v|\frac{3}{2}+u+v\right)}
  {(x_{1,2}^2)^{u+v+2}\,(x_{1,5}^2)^{-u}\,(x_{2,5}^2)^{-v}}
  \label{x0_int2}
\end{equation}
where we have introduced the short notation $\Gamma(z_1|...|z_n)
\equiv\Gamma(z_1)...\Gamma(z_n)$.

We insert this expression back into eq. (\ref{initial}) and perform
the $x_5$--integration. Once again, using Feynman combining we can
write (we neglect factors which are independent of $x_5$)
\begin{eqnarray}
  &&-\epsilon_{\mu\nu\rho}
  \int d^3x_5\,\frac{x_{1,5}^\mu\,x_{2,5}^{\nu}\,x_{4,5}^{\rho}}
  {(x_{1,5}^2)^{-u}\,(x_{2,5}^2)^{-v}\,(x_{3,5}^2)^{1/2}\,(x_{4,5}^2)^{3/2}}=
  \\
  &&-\frac{2\Gamma(2-u-v)}{\Gamma(-u|-v)\pi}
  \int \prod\limits_{i=1}^4dy_i\,\delta(\sum y_i-1)y_1^{-u-1}y_2^{-v-1}y_3^{-1/2}y_4^{1/2}
  \int \frac{d^3x_5\,\epsilon_{\mu\nu\rho}\,x_{1,5}^\mu\,x_{2,5}^{\nu}\,x_{4,5}^{\rho}}{[(x_5-\rho_2)^2+\Omega_2]^{2-u-v}}
  \non
\end{eqnarray}
where  we have defined
\begin{eqnarray}
  && \rho_2^{\mu}=y_1x_1^\mu+y_2x_2^\mu+y_3x_3^\mu+y_4x_4^\mu
  \\
  &&\Omega_2=y_1y_2x_{1,2}^2+y_2y_3x_{2,3}^2+y_3y_4x_{3,4}^2+y_4y_1x_{4,1}^2+y_1y_3x_{1,3}^2+y_2y_4x_{2,4}^2
  \non
\end{eqnarray}
After shifting $x_5^{\mu}\to x_5^\mu+\rho_2^\mu$, we may integrate
over $x_5$ and obtain
\begin{equation}
  \label{int_x6}
  \epsilon_{\mu\nu\rho}\,x_{3,1}^\mu\, x_{3,2}^\nu\, x_{3,4}^\rho\,\frac{2\sqrt{\pi}\,\Gamma(\frac{1}{2}-u-v)}{\Gamma(-u|-v)}
  \int\prod\limits_{i=1}^4dy_i\,\frac{\delta(\sum y_i-1)\,y_1^{-u-1}y_2^{-v-1}y_3^{1/2}y_4^{1/2}}
  {\Omega_2^{1/2-u-v}}
\end{equation}
The first remarkable observation is that this expression is exactly
the Feynman parametrization of a five dimensional scalar square
integral with indices $(-u,-v,3/2,3/2)$. Precisely, we have
\begin{equation}
  \mbox{(\ref{int_x6})}\ \ =
  \epsilon_{\mu\nu\rho}\,x_{3,1}^\mu\,x_{3,2}^\nu\,x_{3,4}^\rho \, \frac{1}{2\pi} \int d^5x_5
  \frac{1}{(x_{1,5}^2)^{-u}\,(x_{2,5}^2)^{-v}\,(x_{3,5}^2)^{3/2}\,(x_{4,5}^2)^{3/2}}
  \label{5dim}
\end{equation}
The identification with a higher dimensional integral is strictly
formal, and should be intended at the level of its
Feynman-parametrized form. In any case, we are dealing with a scalar
integral which depends only on the Lorentz invariants $x_{i,j}^2$ and
these invariants are unambiguously well--defined both in three and
five dimensions.

Collecting all the factors from (\ref{x0_int2}, \ref{5dim}), we are
left with the following expression for the initial integral
\begin{eqnarray}
  \mathcal{I} &=& \frac{2}{\pi^{\frac32}} \, \epsilon_{\mu\nu\rho}\,x_{1,3}^\mu\,x_{2,3}^\nu\,x_{3,4}^\rho
  \int \frac{d^5x_5}{(x_{3,5}^2)^{3/2}\,(x_{4,5}^2)^{3/2}} \times
  \\
  &~&~~~~~~~~~~~~~~~~\int_{-i\infty}^{+i\infty} \frac{du\,dv}{(2\pi i)^2}
  \frac{\Gamma\left(-u|-v|-\frac{1}{2}-u|-\frac{1}{2}-v|2+u+v|\frac{3}{2}+u+v\right)}
  {(x_{1,2}^2)^{u+v+2}\,(x_{1,5}^2)^{-u}\,(x_{2,5}^2)^{-v}}
  \non
\end{eqnarray}
The second remarkable observation is that the MB integral in this
expression can be identified with the MB-representation of a
five dimensional scalar triangle with exponents $(3/2, 3/2,
3/2)$. Therefore, we can write
\begin{equation}
  \mathcal{I}=\frac{1}{4\pi^2} \, \epsilon_{\mu\nu\rho}\,x_{1,3}^\mu\,x_{2,3}^\nu\,x_{3,4}^\rho
  \int d^5x_0\,d^5x_5\,\frac{1}{(x_{0,1}^2)^{3/2}\,(x_{0,2}^2)^{3/2}\,(x_{0,5}^2)^{3/2}\,(x_{3,5}^2)^{3/2}\,(x_{4,5}^2)^{3/2}}
  \label{magic}
\end{equation}
At this point it might seem that we have traded a complicated two-loop
tensor integral in three dimensions with a complicated two-loop scalar
integral in five dimensions. But here comes the magic: We can use the
uniqueness relations applied to the $x_5$--triangle integral.

We recall that for a generic triangle integral in $D$ dimensions with
arbitrary exponents
\begin{equation}
  \mathcal{T}[D;\alpha_1,\alpha_2,\alpha_3;x_{0,3}^2,x_{0,4}^2,x_{3,4}^2]=
  \int \frac{d^D x_5}{(x_{0,5}^{2})^{\alpha_1}\,(x_{3,5}^{2})^{\alpha_2}\,(x_{4,5}^{2})^{\alpha_3}},
\end{equation}
the following identity holds \cite{Usyukina:1994iw}
\begin{eqnarray}
  \mathcal{T}[D;\alpha_1,\alpha_2,\alpha_3; x_{0,3}^2,x_{0,4}^2,x_{3,4}^2] &=&
  \frac{\Gamma(\sum_i \alpha_i- \frac{D}{2})}{\Gamma(D-\sum_i\alpha_i)}
  \, \prod_i \frac{\Gamma(\frac{D}{2}-\alpha_i)}{\Gamma(\alpha_i)} \times \frac{1}{(x_{3,4}^2)^{\alpha_2+\alpha_3-D/2}} \times
  \nonumber\\
  \non \\
  &~&\mathcal{T}\left[D;\sum\alpha_i-\frac{D}{2},\frac{D}{2}-\alpha_3,\frac{D}{2}-\alpha_2;x_{0,3}^2,x_{0,4}^2,x_{3,4}^2 \right]
  \non \\
\end{eqnarray}
Applying this identity to the $x_5$--triangle in (\ref{magic}) where
we identify $D=5$ and $\a_1 = \a_2 =\a_3 = 3/2$, we obtain
\begin{equation}
  \mathcal{I}=\frac{2}{\pi^4} \, \frac{\epsilon_{\mu\nu\rho}\,x_{1,3}^\mu\,x_{2,3}^\nu\,x_{3,4}^\rho}
  { x_{3,4} }
  \int d^5x_0\,d^5x_5\,\frac{1}{(x_{0,1}^2)^{3/2}\,(x_{0,2}^2)^{3/2}\,(x_{0,5}^2)^{2}\,x_{3,5}^2\,x_{4,5}^2}
\end{equation}
The advantage of doing it is that the exponents of the $x_0$ triangle
are now $(3/2,3/2,2)$ and satisfy the uniqueness condition $\a_1 +
\a_2 + \a_3 = D$ in five dimensions. Therefore, we can use the general
result for {\em unique} triangles \cite{Usyukina:1994iw}
\begin{eqnarray}
  && \int \, \frac{d^D x_0}{(x_{0,1}^2)^{\a_1}\,(x_{0,2}^2)^{\a_2}\,(x_{0,5}^2)^{\a_3}} \Big|_{\a_1 + \a_2 + \a_3 = D}
  = \\
  \non \\
  &~&  \qquad \pi^{D/2} \, \prod_i \frac{ \Gamma(D/2-a_i)}{ \Gamma(a_i)}
  \,\frac{1}{ (x_{1,2}^2)^{D/2-\a_3} \, (x_{1,5}^2)^{D/2-\a_2} \, (x_{2,5}^2)^{D/2-\a_1} }
  \non
\end{eqnarray}
and finally write
\begin{equation}
  \mathcal{I}=\frac{8}{\pi^2} \, \frac{\epsilon_{\mu\nu\rho}\,x_{1,3}^\mu\,x_{2,3}^\nu\,x_{3,4}^\rho}
  { x_{1,2}\,  x_{3,4} }
  \int d^5x_5\frac{1}{x_{5,1}^2\,x_{5,2}^2\,x_{5,3}^2\,x_{5,4}^2}
\end{equation}
This concludes the proof of eq. (\ref{5dint}) for $i=1, j=3$. The
generalization of the formula to any $i,j$ is trivial.

\end{document}